\documentclass{ptephy_v1}

\preprintnumber{XXXX-XXXX}

\usepackage{graphics}
\usepackage{url}

\ifdefined \aj \else
\newcommand\aj{{AJ}}%
\newcommand\aaps{{A\&AS}}%
\fi

\begin{document}

\title{Thundercloud Project: Exploring high-energy phenomena in thundercloud and lightning}

\author[1]{Takayuki Yuasa\thanks{Currently, an individual researcher.}}
\affil[1]{High Energy Astrophysics Laboratory, Nishina Center, RIKEN, Wako, Saitama, 351-0106, Japan\email{yuasatakayuki@gmail.com}}

\author[2]{Yuuki Wada}
\author[2]{Teruaki Enoto}
\affil[2]{Extreme Natural Phenomena RIKEN Hakubi Research Team, Cluster of Pioneering Research, RIKEN, Wako, Saitama 351-0198, Japan}

\author[3]{Yoshihiro Furuta}
\affil[3]{Collaborative Laboratories for Advanced Decommissioning Science, Japan Atomic Energy Agency, Futaba, Fukushima 979-1151, Japan}

\author[4]{Harufumi Tsuchiya}
\affil[4]{Nuclear Science and Engineering Center, Japan Atomic Energy Agency, Naka, Ibaraki 319-1195, Japan}

\author[5]{Shohei Hisadomi}
\author[5]{Yuna Tsuji}
\affil[5]{School of Science, Nagoya University, Nagoya, Aichi 464-8601, Japan}

\author[6]{Kazufumi Okuda}
\author[6]{Takahiro Matsumoto}
\affil[6]{Department of Physics, Graduate School of Science, The University of Tokyo, Bunkyo, Tokyo 113-0033, Japan}

\author[7]{Kazuhiro Nakazawa}
\affil[7]{Kobayashi-Maskawa Institute for the Origin of Particles and the Universe, Nagoya University, Nagoya, Aichi 464-8601, Japan}

\author[6,8,9]{Kazuo Makishima}
\affil[8]{High Energy Astrophysics Laboratory, Nishina Center for Accelerator-Based Science, RIKEN, Wako, Saitama 351-0198, Japan}
\affil[9]{Kavli Institute for the Physics and Mathematics of the Universe, The University of Tokyo, Kashiwa, Chiba 277-8583, Japan}

\author[10]{Shoko Miyake}
\affil[10]{Ibaraki College, National Institute of Technology, Hitachinaka, Ibaraki, 312-8508, Japan}

\author[9]{Yuko Ikkatai}

\begin{abstract}%
We designed, developed, and deployed a distributed sensor network aiming at observing high-energy ionizing radiation, primarily gamma rays, from winter thunderclouds and lightning in coastal areas of Japan. Starting in 2015, we have installed, in total, more than 15 units of ground-based detector system in Ishikawa Prefecture and Niigata Prefecture, and accumulated 551~days of observation time in four winter seasons from late 2015 to early 2019. In this period, our system recorded 51 gamma-ray radiation events from thundercloud and lightning. Highlights of science results obtained from this unprecedented amount of data include the discovery of photonuclear reaction in lightning which produces neutrons and positrons along with gamma rays, and deeper insights into the life cycle of a particle-acceleration and gamma-ray-emitting region in a thundercloud. The present paper reviews objective, methodology, and results of our experiment, with a stress on its instrumentation.
\end{abstract}

\subjectindex{H21 Instrumentation for ground observatory, J63 Other topics in geophysics}

\maketitle

\section{Introduction}

Lightning discharges and thunderclouds have been known as electrical phenomena in the atmosphere since the discovery by Benjamin Franklin in 1752. Thanks to recent observational and theoretical studies, they have been also found to be closely associated with high-energy phenomena comprising high-energy photons, electrons, neutrons, etc., and a new academic field ``high-energy atmospheric physics'' has been established. In 1925, Wilson \cite{Wilson_1925} proposed the first idea that strong electric fields in thunderclouds can accelerate $\beta$-particles or electrons of cosmic-ray origin to MeV energies, even in the dense atmosphere. Electrons accelerated in electric fields emit bremsstrahlung photons by colliding with atmospheric nuclei. This Wilson's runaway electron scheme was developed with multiplication processes into relativistic runaway electron avalanches (RREA) by Gurevich et al. \cite{Gurevich_1992}; secondary electrons produced by accelerated electrons via ionization loss processes also become seed electrons and are accelerated in electron fields.

The first reports of high-energy atmospheric phenomena were made by Parks et al. \cite{Parks_1981} and McCarthy and Parks \cite{McCarthy_1985}. They utilized an X-ray counter onboard an F-108 aircraft and detected enhancements of count rates lasting for tens of seconds while flying in thunderclouds. This phenomenon is now called ``gamma-ray glow''. Gamma-ray glows originate from electron acceleration and multiplication in thunderclouds. Their duration ranges from seconds to tens of minutes; their life cycle is thought to be connected to the stability of electric fields inside thunderclouds. So far, gamma-ray glows have been observed by aircrafts \cite{Kelley_2015,Kochkin_2017,Ostgaard_2019}, balloons \cite{Eack_1996}, and mountain-top experiments \cite{Chubenko_2000,Alexeenko_2002,Tsuchiya_2009,Tsuchiya_2012,Torii_2009,Bowers_2019}. When they are detected by ground-based facilities, they are also referred to as thunderstorm ground enhancements (TGEs) \cite{Chilingarian_2010}. In particular, the observatory at Mount Aragats in Armenia has observed the largest number of TGEs by cosmic-ray monitors \cite{Chilingarian_2010,Chilingarian_2012,Chilingarian_2019b}. Gamma-ray glows are sometimes quenched by lightning discharges \cite{McCarthy_1985,Kelley_2015,Kochkin_2017,Alexeenko_2002,Chilingarian_2017,Chilingarian_2019}. This is evidence that electric fields responsible for gamma-ray glows can be destroyed by lightning currents.

Besides airborne and mountain-top observations of gamma-ray glows, experiments during winter thunderstorms in Japan are of great importance. In coastal areas facing the Sea of Japan, northern seasonal winds blow and provide heavy snow with lightning discharges. These winter thunderstorms in Japan are distinctive comparing to typical thunderstorms, in particular cloud bases. While typical summer thunderstorms develop above an altitude of 3-km or higher, winter thunderclouds in Japan have a cloud base of lower than 1~km \cite{Kitagawa_1994}. Gamma-ray photons are absorbed in the atmosphere typically within 1~km. Therefore, we need in-situ measurements by airborne detectors or getting closer to thunderclouds by putting detectors on mountain tops, to observe gamma rays from summer thunderstorms. On the other hand, winter thunderstorms allow us to observe high-energy atmospheric phenomena at sea level. Torii et al. \cite{Torii_2002} reported gamma-ray glows lasting for $\sim$1~minute during winter thunderstorms for the first time, recorded by dosimeters installed at a nuclear power facility in a coastal area of the Sea of Japan. Another measurement with multiple dosimeters succeeded in tracking a gamma-ray glow moving with a thundercloud and ambient wind flow \cite{Torii_2011}.

Another important class of high-energy atmospheric phenomena is ``terrestrial gamma-ray flash'' (TGF). TGFs are transient emission coinciding with lightning discharges. Their energy spectrum extends up to $>20$~MeV \cite{Smith_2005,Tavani_2011}, and their duration is typically several hundreds of microseconds \cite{Foley_2014}. Since their discovery by Compton Gamma-Ray Observatory \cite{Fishman_1994}, they have been routinely detected by in-orbit experiments such as RHESSI \cite{Smith_2005}, AGILE \cite{Tavani_2011,Marisaldi_2010}, Fermi \cite{Briggs_2010,Mailyan_2016}, and ASIM \cite{Neubert_2019b,Ostgaard_2019b}. TGFs are thought to be produced by $10^{16}$--$10^{19}$ energetic electrons above 1~MeV \cite{Dwyer_2005c,Mailyan_2016}. While several models have been proposed \cite{Celestin_2011,Celestin_2015,Moss_2006,Dwyer_2007,Dwyer_2012a}, the mechanism to produce such an enormous number of energetic electrons is still in debate. TGFs detected from space are upward-going, namely emitted from thunderclouds into space. More recently, downward-going ones called ``downward TGFs'' have been detected by ground-based experiments \cite{Dwyer_2004,Hare_2016,Tran_2015,Abbasi_2018,Ringuette_2013}.

Motivated by the initial findings in 1990s and early 2000s, we launched the Gamma-Ray Observation of Winter Thunderclouds (GROWTH) experiment in 2006. The GROWTH experiment is a ground-based measurement of gamma rays and high-energy particles aiming at detecting and exploring high-energy atmospheric phenomena during winter thunderstorms in Japan. The experiment started with a suite of gamma-ray and electron detectors installed at Kashiwazaki-Kariwa Nuclear Power Station of Tokyo Electric Power Holdings in Niigata Prefecture, Japan. The power station faces the Sea of Japan, and frequently encounters lightning discharges in winter seasons. Tsuchiya et al. \cite{Tsuchiya_2007} reported the first detection of a gamma-ray glow lasting for $\sim$40~sec in the Kashiwazaki-Kariwa site. Its energy spectrum, a continuum extending up to 10~MeV originating from bremsstrahlung of electrons, suggested that a thundercloud continuously accelerated electrons to 10~MeV or higher energy. Combining Monte-Carlo simulations, glows in the site were found to originate at an altitude of$<$1~km \cite{Tsuchiya_2011}. Tsuchiya et al. \cite{Tsuchiya_2013} reported a glow abruptly terminated with a lightning discharges. The energy spectrum of the glow gradually became hard, i.e. the ratio of $>$10~MeV to 3--10~MeV photons was increasing as the lightning discharge was drawing near. Umemoto et al. \cite{Umemoto_2016} reported an enigmatic enhancement of electron-positron annihilation gamma rays after a lightning discharge.

During the first decade of the GROWTH experiment in 2006--2015, one or two observation points were maintained at the power station. However, the sparse distribution of detectors is not sufficient to delve deeper into the nature of gamma-ray glows such as on-ground distribution of particles and the life cycle of their acceleration site, i.e. how particle acceleration is initiated, develops, and comes to an end. Therefore, we launched a new campaign of the GROWTH experiment with multiple gamma-ray detectors and observation sites, called ``Thundercloud Project'' in 2015. The initial scientific results of the campaign have been already reported as journal publications \cite{Enoto_2017,Wada_2018,Wada_2019_commphys,Wada_2019_prl,Wada_2020}. In this paper, we describe the design and the performance of our gamma-ray detector system followed by details of completed observation campaigns. Highlights of scientific achievements obtained based on data from these observation campaigns are also summarized.

Throughout the paper, gamma-ray glow (gamma-ray emission from the thunder cloud, typically lasting for a few minutes) and short-duration gamma-ray burst caused by a downward TGF (lasting for a fraction of second) are collectively called thundercloud radiation bursts (TRBs). When we put a stress on the time scale of gamma-ray burst events from the observational point of view, we also call minute-lasting gamma-ray glow as ``long-duration gamma-ray burst''.

\section{Experiment setup: gamma-ray detector system}

At high level, our detector system is a conventional photon-counting gamma-ray spectrometer based on scintillation crystal and photo-multiplier tube (PMT). For realizing a distributed observation network of TRBs, we set miniaturization of an entire system as a primary design goal to allow easy handling and deployment in rooftop/outdoor environments.
Keeping the cost of the system as low as possible is also essential because otherwise the number of detector systems manufactured would not be large due to the tight research budget and the scale of the observation network would be limited.
In addition, we deploy the detector system to multiple locations (distances between detectors varying from a few to hundreds of km), frequent on-site maintenance is not an option, and remote-monitoring and remote-control capabilities are indispensable.

\begin{figure}[htb]
\begin{center}
\includegraphics[width=1\hsize]{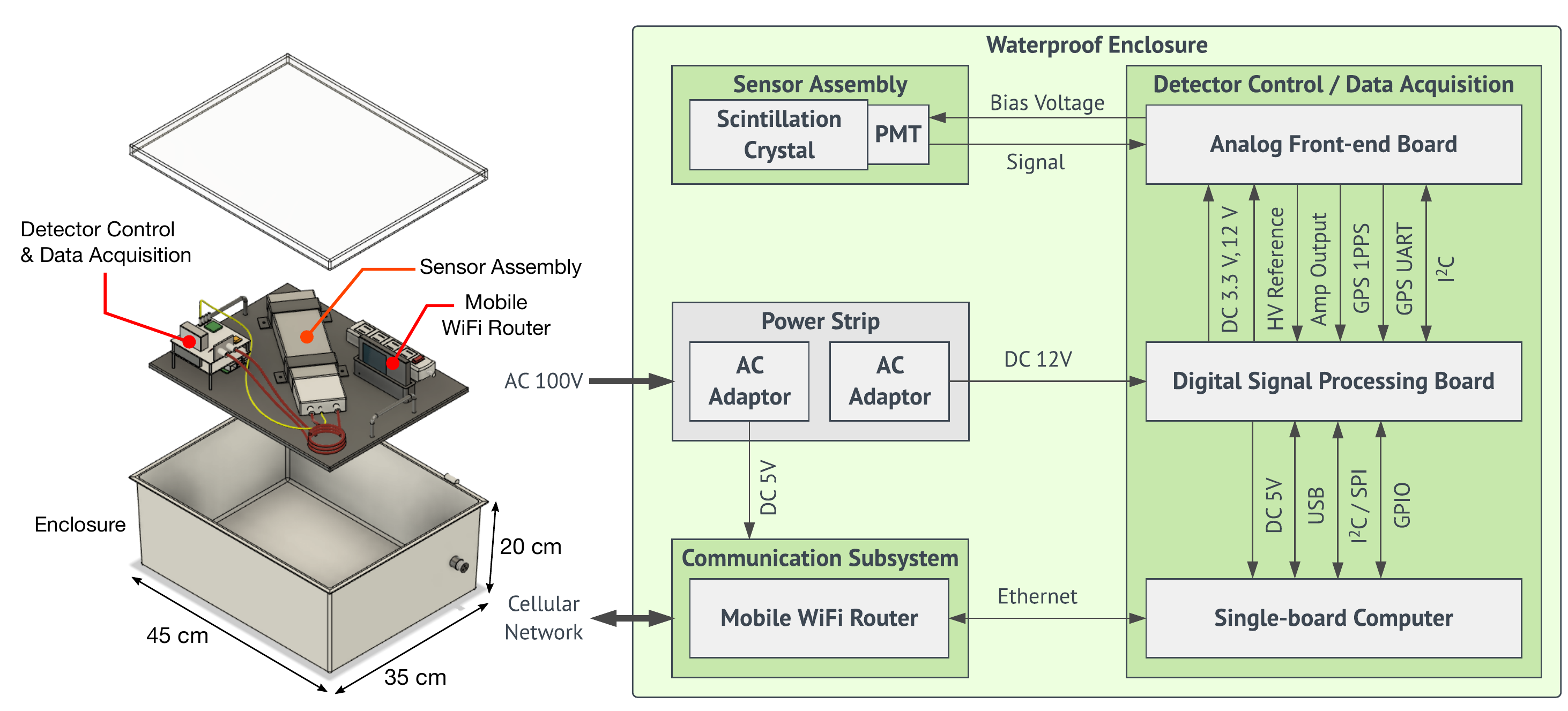}
\end{center}
\caption{Exploded view of the CAD drawing (left) and system block diagram (right) of the gamma-ray detector system. The size of the detector is 35~cm (depth)$\times$45~cm (width)$\times$20~cm (height).}
\label{fig:block_diagram}
\end{figure}

Figure~\ref{fig:block_diagram} shows a high-level block diagram of the detector system, which consists of 1) scintillation crystal viewed with photo-multiplier tube (hereafter sensor assembly), 2) Detector-control and data-acquisition electronics subsystem (hereafter DAQ subsystem), 3) telecommunication subsystem, and 4) mechanical support structure and waterproof enclosure. The detector is supplied with AC 100~V from the commercial power line, and a switching regulator generates DC voltages (12~V and 5~V) required by the electronics subsystem and the telecommunication subsystem. The telecommunication subsystem provides internet connectivity via a cellular network, and is used for telemetry transmission from the detector system, and remote-login to a computer in the DAQ subsystem via secure shell (ssh). Typical power consumption of the entire system is about 7~W.

In the following subsections, detailed specifications of individual subsystems are described.

\subsection{Sensor assembly}

For detailed temporal and spectral analyses, it is critically important to detect gamma rays from thundercloud and lightning at as high photon counts as possible. The only way to achieve this is to make the effective area of a sensor larger and to select sensor materials with high stopping power against gamma rays with energies of MeV to a few tens of MeV.

Bismuth germanite (Bi$_{4}$Ge$_{3}$O$_{12}$; hereafter BGO) scintillation crystal is one of optimal crystals in the thundercloud gamma-ray observation due to its high stopping power and environmental durability (no deliquescence). In the pilot observation campaign in 2015, we have employed cylindrical BGO crystals each with a diameter of 7.62~cm and a height of 7.62~cm. The standard BGO crystals that we used in the regular observation campaign since 2016 have dimensions of 25~cm~$\times$~8~cm~$\times$~2.5~cm. One crystal is viewed with two HAMAMATSU R1924A PMTs, and outputs from the two PMTs are combined in the analog stage, and then amplified and digitized as a single signal. Each set of a crystal and two PMTs are enclosed in a 2-mm-thick aluminum case. We have used 15 of this BGO-based sensor assemblies since 2016.

During our detector development, low-cost Thallium-doped Cesium Iodide crystals, or CsI (Tl) for short, that were extracted from a terminated accelerator experiment project became available, and we have purchased a dozen of 30~cm~$\times$~5~cm~$\times$~5~cm crystals. The effective area of the CsI-based sensor assembly is slightly smaller than that of BGO-based ones, but helped to expand our observation network at a moderate increment of the manufacturing cost.

\begin{figure}[htb]
\begin{center}
\includegraphics[width=0.7\hsize]{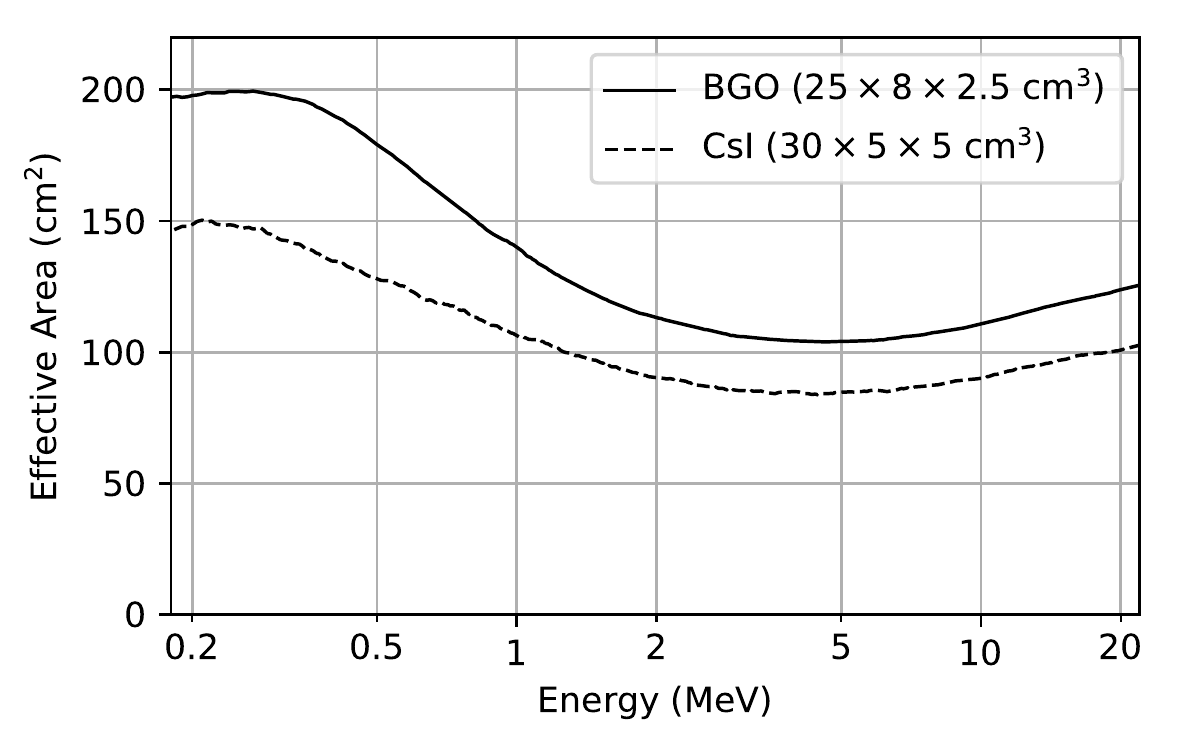}
\end{center}
\caption{Effective area of each scintillation crystal calculated by a Monte-Carlo simulation. Uniformly-distributed gamma rays arriving from the direction of the normal of the detection surface (25~cm~$\times$~8~cm face of BGO and 30~cm~$\times$~5~cm face of CsI) are assumed in the simulation. Photo absorption, Compton scattering, and electron-positron pair creation are the physical processes involved in the simulation, and interactions that deposited energies larger than 40~keV in the crystal were considered detectable.}
\label{fig:effective_area}
\end{figure}

Figure~\ref{fig:effective_area} illustrates the effective area of each scintillation crystal over gamma-ray energies 0.2--20~MeV, which is a typical energy range our detectors observe.
Table \ref{tab:energy_scale} summarizes the energy resolution of the BGO and CsI crystals, measured via the laboratory calibration (0.662~MeV from $^{137}$Cs isotope) and using the environmental background signal (1.46~MeV and 2.61~MeV from $^{40}$K and $^{208}$Tl, respectively).

\begin{table}[!h]
\caption{Typical energy resolution of the detector.}
\label{tab:energy_scale}
\centering
\begin{tabular}{cccc}
\hline
Crystal & \multicolumn{3}{c}{Resolution} \\
        & 0.662~MeV  &  1.46~MeV &  2.61~MeV \\
\hline
BGO     & 19\% & 12\% & 9\% \\
CsI(Tl) & 12\% &  9\% & 7\% \\
\hline
\end{tabular}
\end{table}

\subsection{Detector-control and data acquisition (DAQ) subsystem}

We developed a data-acquisition and detector-control system based on 1) an analog front-end board, 2) digital signal-processing (DSP) board, and 3) commercial-off-the-shelf single-board computer Raspberry Pi.
The analog front-end board is a custom board designed by our group specifically for the present experiment.
The DSP board is a general purpose Field Programmable Gate Array (FPGA) board with 4-ch waveform-sampling Analog-to-Digital Converters (ADCs). We developed the FPGA/ADC board in collaboration with Shimafuji Electric, primarily for our experiment but also aiming for broader applications in other projects.

\begin{figure}[htb]
\begin{center}
\includegraphics[width=0.6\hsize]{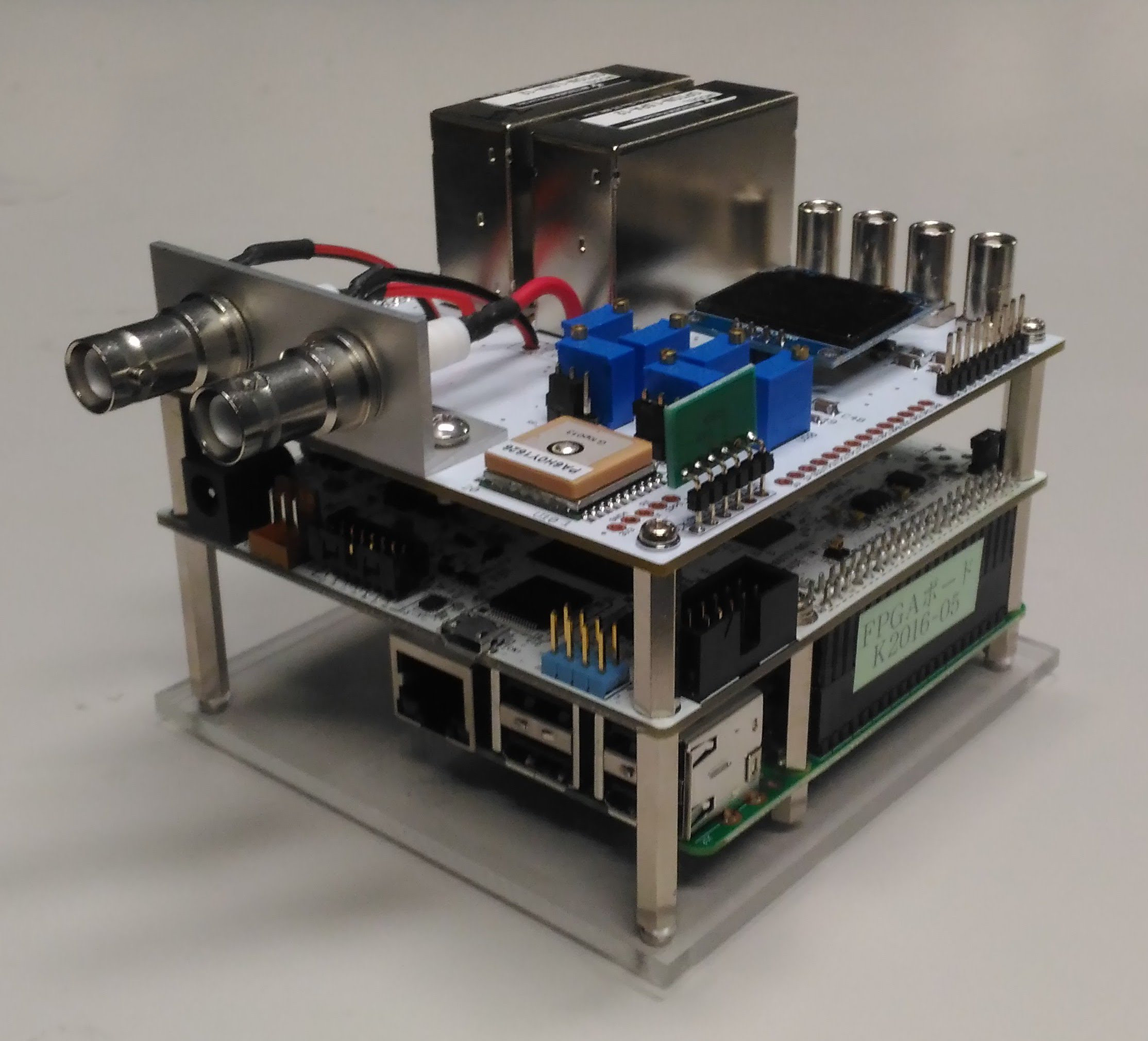}
\end{center}
\caption{Outlook of the manufactured DAQ subsystem. From top to bottom, the front-end analog signal processing board, the FPGA/ADC board, and the Raspberry Pi computer can be seen. The size of the stacked subsystem is about 10~cm (depth)$\times$10~cm (width)$\times$10~cm (height).}
\label{fig:daq}
\end{figure}

As shown in Fig.~\ref{fig:daq}, these boards are vertically stacked using 2.54-mm-pitch board-to-board connectors, forming a standalone data acquisition system within a cube of 10$\times$10$\times$10~cm$^3$, excluding protruding high-voltage power supply connectors. This design was chosen to save footprint of the system, and also to reduce required cabling during fabrication and integration at each observation site. Though the entire DAQ system is compact, it fully implements analog and digital signal processing required to function as a gamma-ray spectrometer and autonomously collect data for several months.
Since we consider this miniaturized DAQ system as one of key enablers of our multi-point observation campaign, the design of the system is detailed in the following paragraphs. The high-level technical specification of the system is also summarized in Table \ref{tab:daq_spec}.

\begin{table}[!h]
\caption{Specification of the DAQ system.}
\label{tab:daq_spec}
\centering
\begin{tabular}{cl}
\hline
\multicolumn{2}{c}{Analog Front-end Board}\\
\hline
Function & Specification \\
\hline
Amplifier & Custom design. 4~channels. Pass band $\sim$2~$\mu$s. \\
HVPS & OPTON-1.5PA (Matsusada)$\times$2. Up to 1.5~kV.\\
GPS & FGPMMOPA6H.\\
    & On-chip patch antenna or external antenna via SMA connector.\\
OLED display & 128$\times$64 pixels. 0.9~inch. I2C.\\
Env. Sensor & Temperature, humidity, pressure with BME280 (Bosch Sensortec). I2C.\\
Stack connector & 20$\times$2 pin header.\\
\hline
 & \\
\hline
\multicolumn{2}{c}{Digital Signal Processing Board}\\
\hline
Function & Specification\\
\hline
FPGA & Artix-7 XC7A35T-1FTG256C (Xilinx). System clock 50~MHz.\\
GPIO & HVPS output enable, GPS 1PPS and NMEA data reception.\\
     & 6 GPIO pins to Raspberry Pi.\\
ADC  & AD9231BCPZ-65 (Analog Devices)$\times$2.\\
     & 4-ch, 12-bit, 50-Msps sampling. Input range $\pm$5~V.\\
Slow ADC & MCP3208-BI/SL (Microchip). 4-ch, 12-bit sampling. SPI.\\
         & 4-ch left for user.\\
Slow DAC & MCP4822-E/MS (Microchip). 2-ch, 12-bit sampling. SPI.\\
         & Used for HVPS reference voltage.\\
USB interface & FT2232HL (FTDI Chip). USB Micro-B connector.\\
Temp. sensor & LM60BIM3 (Texas Instruments).\\
             & Provides FPGA and DC/DC converter temperatures.\\
Current sensor & LT6106HS5 (Linear Technology). I2C.\\
               & Provides 12~V, 5~V, 3.3~V current consumption.\\
Stack connector & 20$\times$2 pin socket for Analog Front-end Board.\\
 & 20$\times$2 pin header for Raspberry Pi.\\
Power & 12~V via 2.1~mm jack.\\
 & $\sim7$~W power consumption in the nominal observation mode.\\
Dimension & 9.5$\times$9.5$\times$2.9~cm$^3$.\\
\hline
 &  \\
\hline
\multicolumn{2}{c}{Raspberry Pi 3}\\
\hline
Function & Specification\\
CPU & Quad-core 1.2-GHz ARM Cortex-A53.\\
RAM & 1~Gigabytes.\\
Storage & 32~Gigabytes, Class 10 SD card.\\
USB & 4~ports.\\
Ethernet & 100~Base Ethernet.\\
\hline
\end{tabular}
\end{table}

\subsubsection{Analog front-end board}

The analog front-end board carriers high-voltage power supply (HVPS) modules, amplifier chains, a Global Positioning System (GPS) receiver, and an organic light-emitting-diode (OLED) display.
The board also implements a combined temperature, pressure, and humidity sensor BME-280 for providing house-keeping information.

We selected the OPTON-1.5PA HVPS module from Matsusada Precision as our system, because of its small footprint and volume (44$\times$30$\times$16~mm$^3$).
The board can carry up to two HVPS modules and high-voltage outputs from the modules are routed to two Safe High Voltage (SHV) connectors.
The reference voltage signals of the HVPS modules are connected to a 2-ch 12-bit digital-to-analog converter MCP4822-E/MS on the digital signal processing board, so that output voltages can be flexibly controlled from software on Raspberry Pi via Serial Peripheral Interface (SPI).

\begin{figure}[htb]
\begin{center}
\begin{minipage}{0.45\hsize}
\includegraphics[width=\hsize]{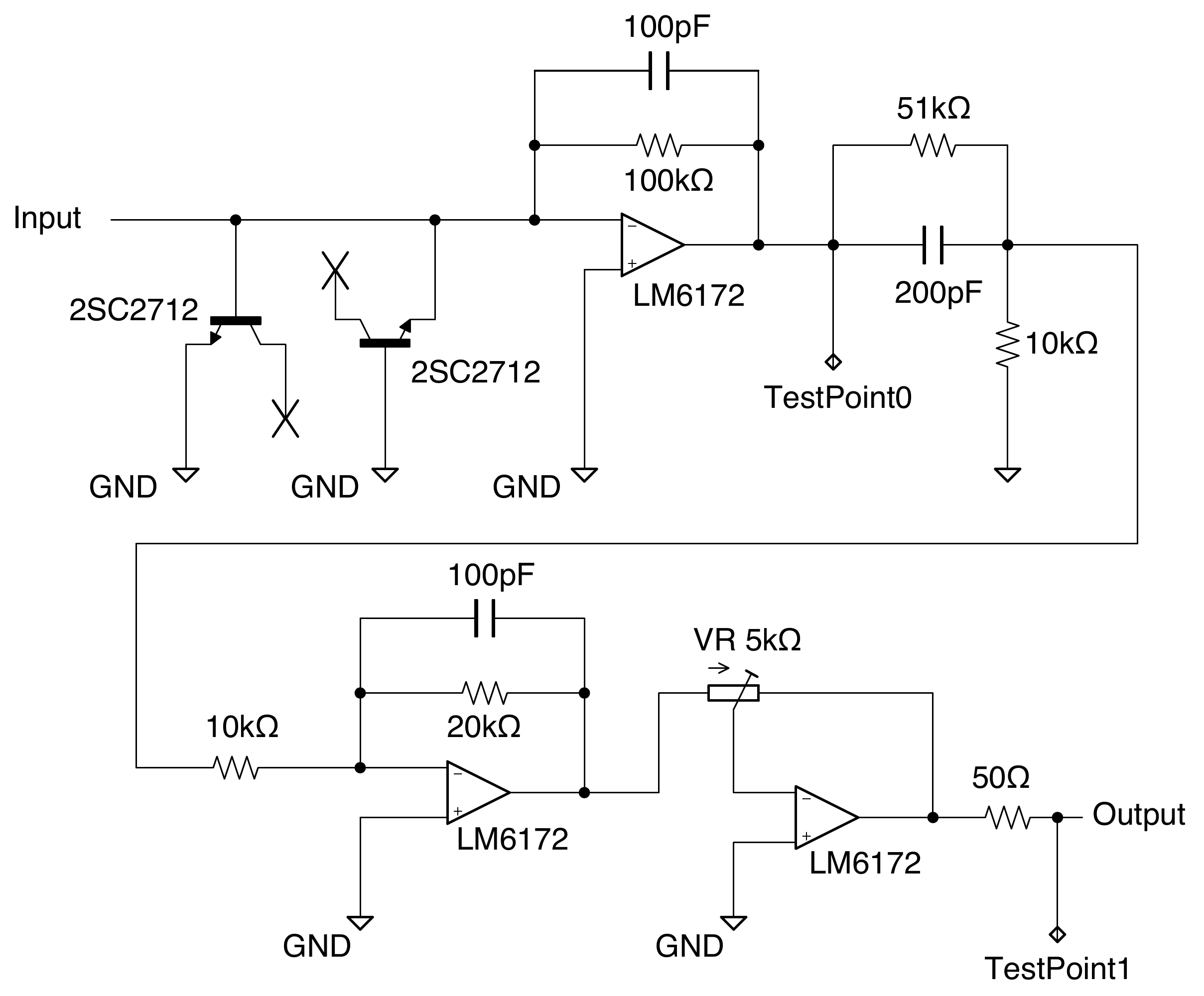}
\end{minipage}
\begin{minipage}{0.45\hsize}
\includegraphics[width=\hsize]{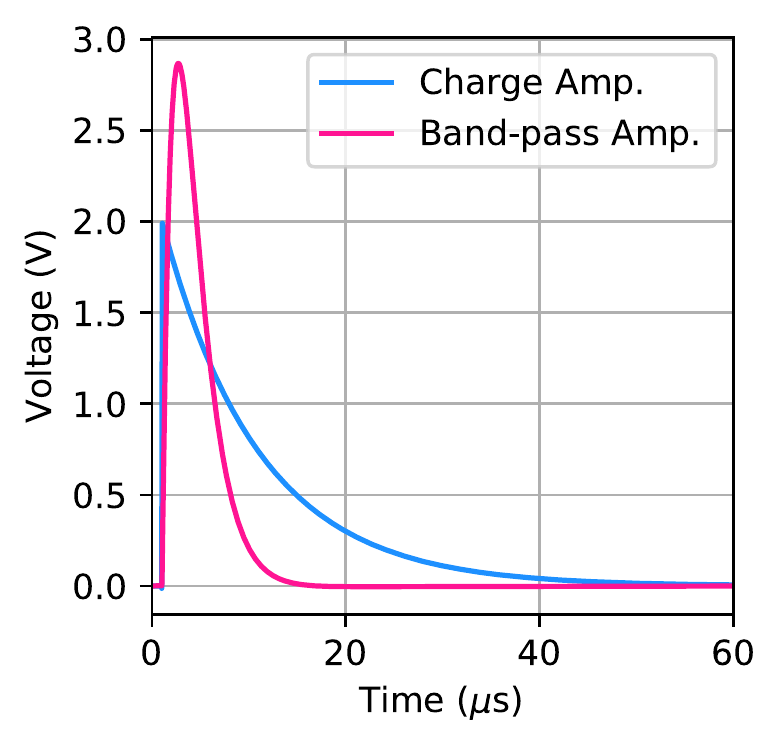}
\end{minipage}
\end{center}
\caption{Left: A circuit diagram of the analog amplifier chain. Bypass capacitors used for operational amplifiers are not shown in this diagram for simplicity. Right: SPICE-simulated pulse shapes. Blue and pink voltage waveforms are measured at Test Points 0 and 1 of the circuit diagram, respectively, against a typical charge supplied by a BGO+PMT assembly.}
\label{fig:amp}
\end{figure}

The amplifier chain consists of a simple charge-integration amplifier that converts a charge output of the PMTs to a voltage signal, followed by a differentiator-integrator band-pass filter and a linear amplifier. Figure~\ref{fig:amp} shows a circuit diagram of the chain. Four copies of the same amplifier chains are implemented. When a pulse of charge with a decay time of $\sim$300~ns is fed from the sensor assembly (BGO and PMT) to the first-stage charge-integration amplifier, an output pulse from the band-pass-filter amplifier should look like a uni-polar pulse with a $\sim$1~$\mu$s rise and $\sim$4~$\mu$s fall timescales as shown in the right panel of the figure.
These timescales are sufficiently slow compared with the sampling frequency of the waveform-sampling ADC on the digital signal processing board (see the next section), and therefore, the peak pulse height, which is proportional to the energy deposit in the scintillation crystal, can be accurately measured.

The OLED display is connected to the Inter-integrated Circuit (I2C) bus of Raspberry Pi via board-to-board connectors, and is controlled by a simple Python program running on Raspberry Pi that prints a status and parameters of the system, such as observation mode, high-voltage output values, an Internet Protocol (IP) address, and so on. Although the size of the display is small ($\sim$1~inch diagonal) and the resolution is very limited (128$\times$64 pixels), the display turned out to be very helpful in understanding the state of the DAQ system during in particular outdoor deployment works thanks to its high visibility.

\subsubsection{Digital Signal Processing (DSP) board}

\begin{figure}[htb]
\begin{center}
\includegraphics[width=0.97\hsize]{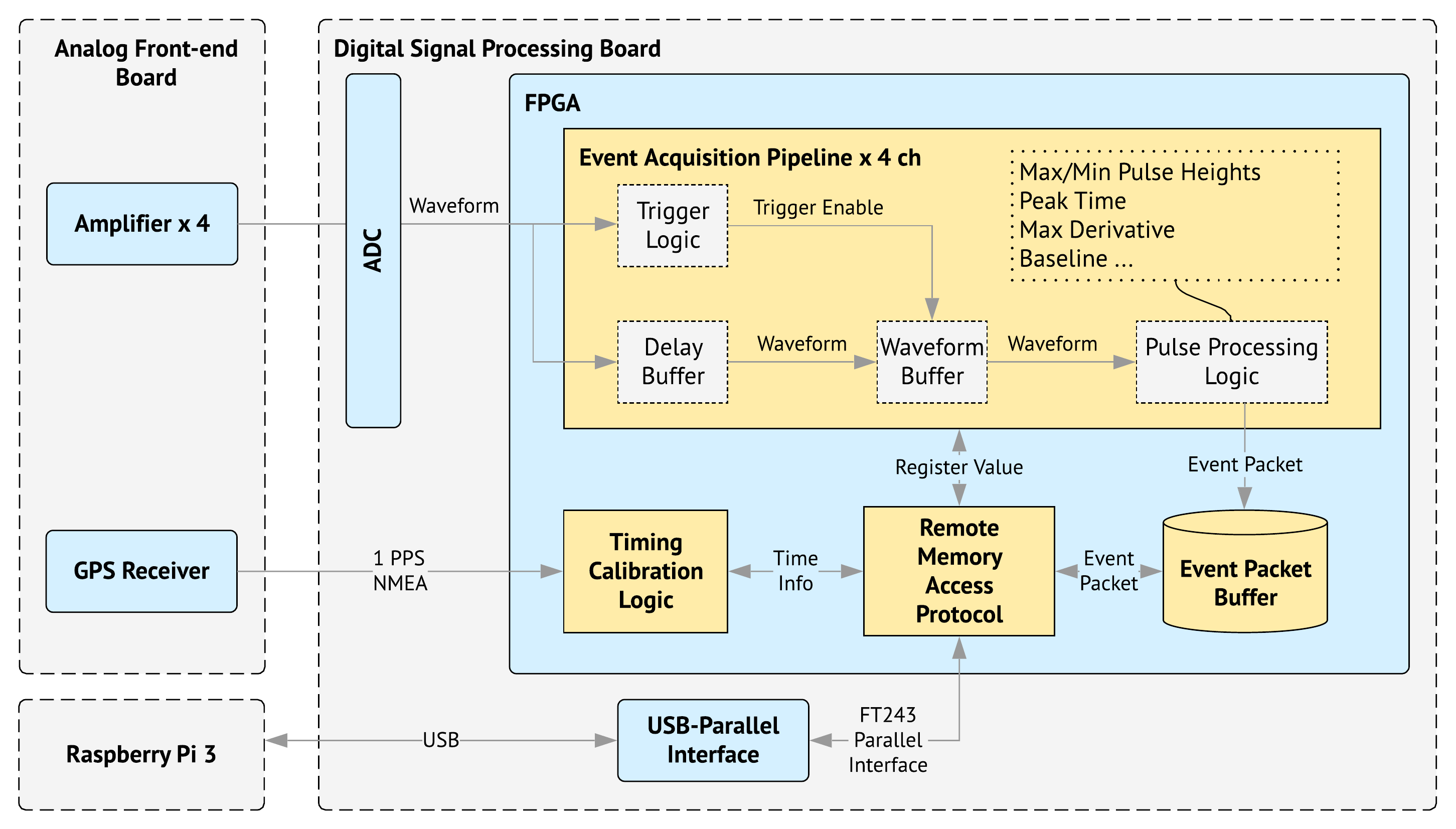}
\end{center}
\caption{High-level block diagram of the FPGA logic.}
\label{fig:fpga_logic}
\end{figure}

The DSP board is a custom-made digitizer consisting of a Xilinx Artix-7 FPGA (XC7A35T-1FTG256C) and two dual 12-bit ADC (Analog Devices AD9231BCPZ-65) that operate at 50~million samples/s (Msps), temperature and current sensors, USB interface (FTDI Chip FT2232H), slow ADC/DAC, and DC/DC converters.
A custom hardware logic that collects timing and pulse height of gamma-ray signals in a self-trigger mode, was developed in Hardware Description Language (HDL) and programmed to FPGA.
Figure~\ref{fig:fpga_logic} illustrates the high-level block diagram of the FPGA logic.
Once the input voltage exceeds the trigger threshold, a predefined number of ADC values are recorded as a "waveform" in Waveform Buffer (typically covering 10~$\mu$s since the trigger), and various properties of the waveform is then computed (maximum pulse height as well as supplementary data such as ADC values of the first/last/minimum pulse heights, sample index of the maximum pulse height, and maximum derivative of waveform values). The maximum pulse height is converted to energy deposit in the crystal in the post processing, and the supplementary data can be used to verify the normal operation of the electronics (PMT, amplifier, ADC) when necessary (see e.g. \cite{Enoto_2017} for use of the supplementary data in addition to the pulse height data). The derived properties are then packed into a certain data packet structure, and stored in Event Packet, and then read by the data acquisition program running on the single-board computer via USB. The source code is publicly available on our project's online repository\footnote{https://github.com/growth-team/}.

The analog front-end board and Raspberry Pi are connected to the DSP board via two 20$\times$2-pin 2.54-mm pitch connectors placed near two edges of the board. The PMT output signal amplified by the analog front-end board and the 1~PPS/NMEA output from the GPS module are routed to ADC and FPGA, respectively, via the connector.
The I2C signal, HVPS reference voltage from the slow DAC, and output enable signals are also passed through the connector from the DSP board to the analog front-end board.

The I2C from the analog front-end board and the SPI communication signals of the slow ADC/DAC, and the HVPS output status control signals are connected to Raspberry Pi via the other 20$\times$2 pin header connector.

The DSP board was manufactured by Shimafuji Electric, and is available for customers as a general-purpose waveform-sampling ADC/FPGA board.

\subsubsection{Single-board computer}

We selected the Raspberry Pi single-board computer as our platform to run software programs that control the HVPS output mode and output voltages, collect gamma-ray event data from the DSP board, read housekeeping data from the house-keeping sensors, and transmit the house-keeping data and status information to the internet. Primary reasons of the selection include its small size (85$\times$56$\times$17~mm$^3$), low price ($<$US\$100 including a power adapter and an SD card), and sufficiently high performance with a quad-core ARM Cortex-A53 processor running at 1.4~GHz and 1~gigabytes (GB) main memory.

The data collection program is the only performance-critical program as the processing speed of it limits the number of gamma-ray events that the entire detector can record, and is written in C++. When the detector system is powered on, the program configures the FPGA logic on the DSP board; for example, it sets enabled ADC channels, trigger threshold values, and the number of waveform samples to be recorded per trigger. When a data collection is started, the program continuously reads the data stored on the Event Packet Buffer of the DSP board, and save the read data to a file as an event list; supported output file formats are CERN/ROOT \cite{BrunRademakers_1997} and FITS \cite{Wells_1981}. 
Ruby and Python are used for other non-performance critical programs to expedite the development by leveraging existing software libraries provided in the ecosystem of these scripting languages such as an OLED display controller library, a digital general-purpose input/output library, and so on.

The process monitoring framework God\footnote{http://godrb.com/} was used to run these programs as resident processes (so-called daemons) after power on.
Configurations of the programs, such as HVPS output voltages, trigger threshold, enabled ADC channels, and data collection mode (i.e. whether to start data collection and HVPS output automatically after boot) are stored in a file on the non-volatile memory (micro SD card) along with the programs and the Linux operating system.

Data collected by the programs, for example, gamma-ray event-list data, house-keeping data, are also stored on the micro SD card. An external flash memory disk connected via Universal Serial Bus (USB) was also used as a back-up storage, and the data are regularly copied from the micro SD card to the external disk.

\subsection{Telecommunication subsystem}

Raspberry Pi in the DAQ subsystem was connected via Ethernet to a mobile WiFi router (Aterm MR04LN from NEC) that is connected to the internet over a cellular network. Due to the stringent monthly data limitation (1~GB per 1~months) of the cellular plan that was allowed by the research grant expenditure regulation, it was infeasible to transfer all the gamma-ray event list data that amount $\sim$5--10~GB per month to a remote data-storage server, and therefore the connectivity was primarily used to transmit the low-data-rate telemetry sent every 300~s and a digest report of gamma-ray data such as binned count-rate histories and time-integrated energy spectra.

\begin{figure}[htb]
\begin{center}
\includegraphics[width=0.7\hsize]{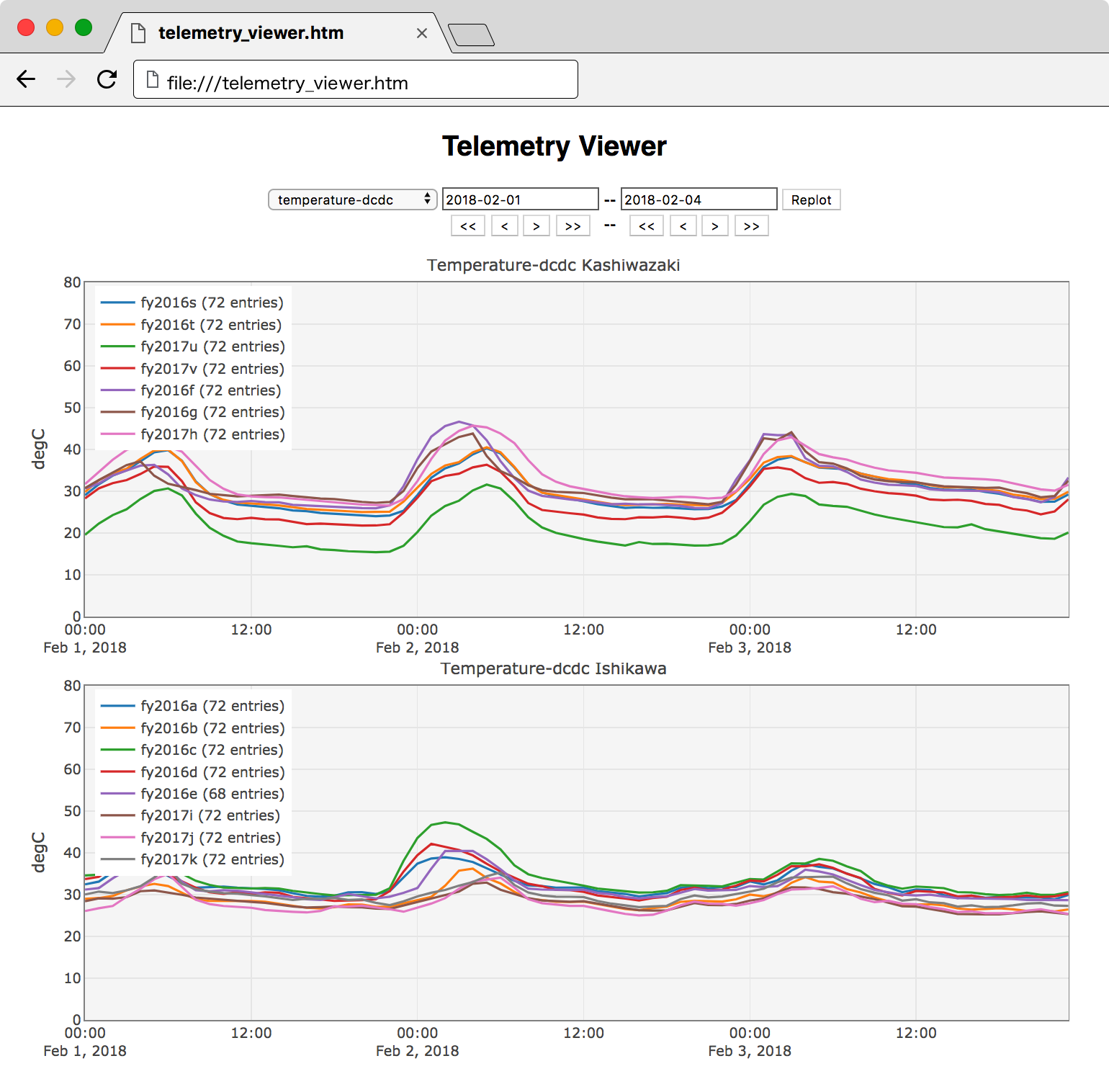}
\end{center}
\caption{Web-browser-based telemetry visualization tool. This example screen shot shows temperature of the DC/DC module on the FPGA/ADC board over a 4-day period in February 2018. The top and the bottom panels are for detector systems deployed in Niigata Prefecture and Ishikawa Prefecture, respectively. }
\label{fig:telemetry_viewer}
\end{figure}

The telemetry data were sent to a cloud-based data base, and this allowed centralized monitoring of the status of the distributed detector systems using a web-browser-based data visualization tool. Figure~\ref{fig:telemetry_viewer} shows an example screen shot of the temperature telemetry. During observation campaigns, occasional stoppages of Raspberry Pi, which is thought to arise from instantaneous AC power failure, were noticed as absence of telemetry data, enabling prompt actions such as power cycling (reboot) by local support personnel.

The digest report of gamma-ray data were also useful in rapidly identifying gamma-ray enhancement events originating from thundercloud and/or lightning; when an enhancement event candidate is noticed, we remote-logged in to Raspberry Pi via ssh and manually transferred a limited number of data files for in-depth analyses.

Having "bi-directional" connectivity to individual detector systems thus helped a day-to-day operation during observation campaigns and also contributed to reduce latency between observation and data analysis, and to expedite publication of the data.
If we were unable to retrieve data remotely, the time/human resource/financial costs of frequent data retrieval, for example, once per month, would have been impractically expensive. Therefore, we consider that the cost of installing a mobile WiFi router ($\sim$US\$100) and purchasing cellular data plan for each detector system ($\sim$US\$10 per month) have been well paid off.

\subsection{Mechanical structure}

\begin{figure}[htb]
\begin{center}
\includegraphics[width=0.8\hsize]{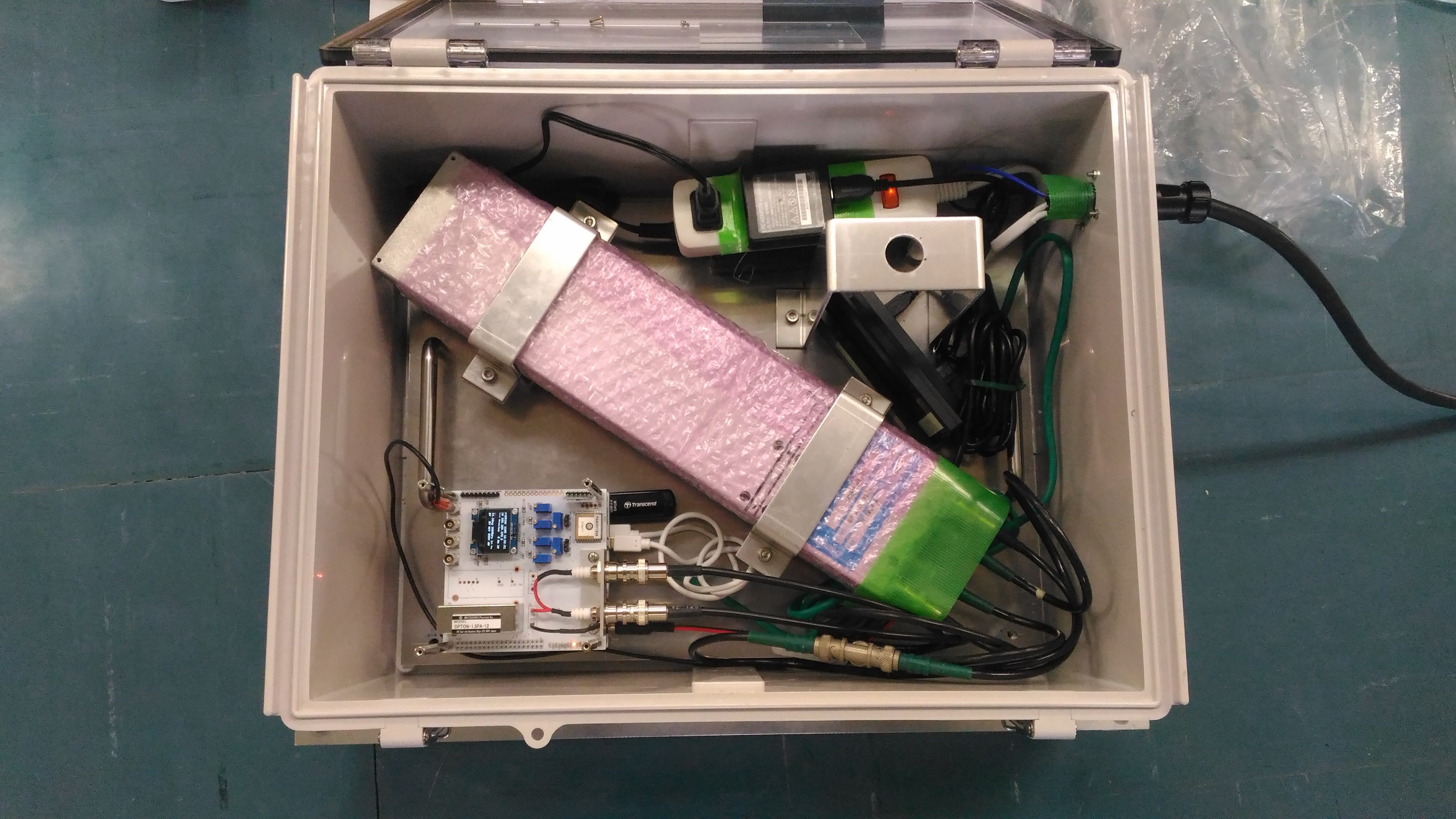}
\end{center}
\caption{Interior of the water-proof enclosure. The BGO-based sensor assembly wrapped with a pink bubble wrap and the DAQ subsystem are located in the center and in the left bottom corner, respectively. A small black module adjacent to the sensor assembly is a mobile WiFi router. A power strip and AC adapters for the DAQ subsystem and the mobile WiFi router are placed in the top right corner.}
\label{fig:enclosure}
\end{figure}

For operating detectors in outdoor environments where snow and sea wind are the norm, we selected the Takachi Electronics Enclosure's water-proof and dust-tight plastic enclosure family BCAR as containers of our detector systems. The dimensions of a standard enclosure we used are 35$\times$45$\times$20~cm$^3$.
A water-proof power connector was attached to one side of an enclosure to pass through AC 100~V power.
When integrating an entire detector system, a sensor assembly (or multiple of them, depending on configuration), a DAQ subsystem, and a telecommunication subsystem were screw-mounted on an aluminum base plate for ruggedization, and then the base plate was screw-mounted to the base of a water-proof enclosure, as shown in Fig.~\ref{fig:enclosure}.

\section{Calibration and offline data analysis}

After each observation campaign in winter, data stored on the detector are retrieved from each detector system, and the energy and the timing calibrations are applied as detailed in \S\ref{sec:energy_cal} and \S\ref{sec:timing_cal}.
Based on the energy- and time-calibrated data, gamma-ray enhancement events, both long- and short-duration ones, are searched using a count-history-based algorithm that is described in \S\ref{sec:search_algorithm}.

\subsection{Energy scale calibration}\label{sec:energy_cal}

During outdoor observations, the energy scale changes over time as ambient temperature and temperature of the scintillation crystal vary. Instead of actively compensating this change by for example dynamically adjusting the PMT or the analog amplifier gain, we let the detector operate at the pre-determined fixed gain, and corrected energy scale in the offline analysis. The correction was made by fitting the prominent gamma-ray lines seen in the environmental background radiation spectrum such as lines at 1.46~MeV ($^{40}$K), and 2.61~MeV ($^{208}$Tl) in the ADC channel space, and by constructing the best-fit linear function which returns energy in MeV for a given ADC channel.

\begin{figure}[htb]
\begin{center}
\includegraphics[width=0.8\hsize]{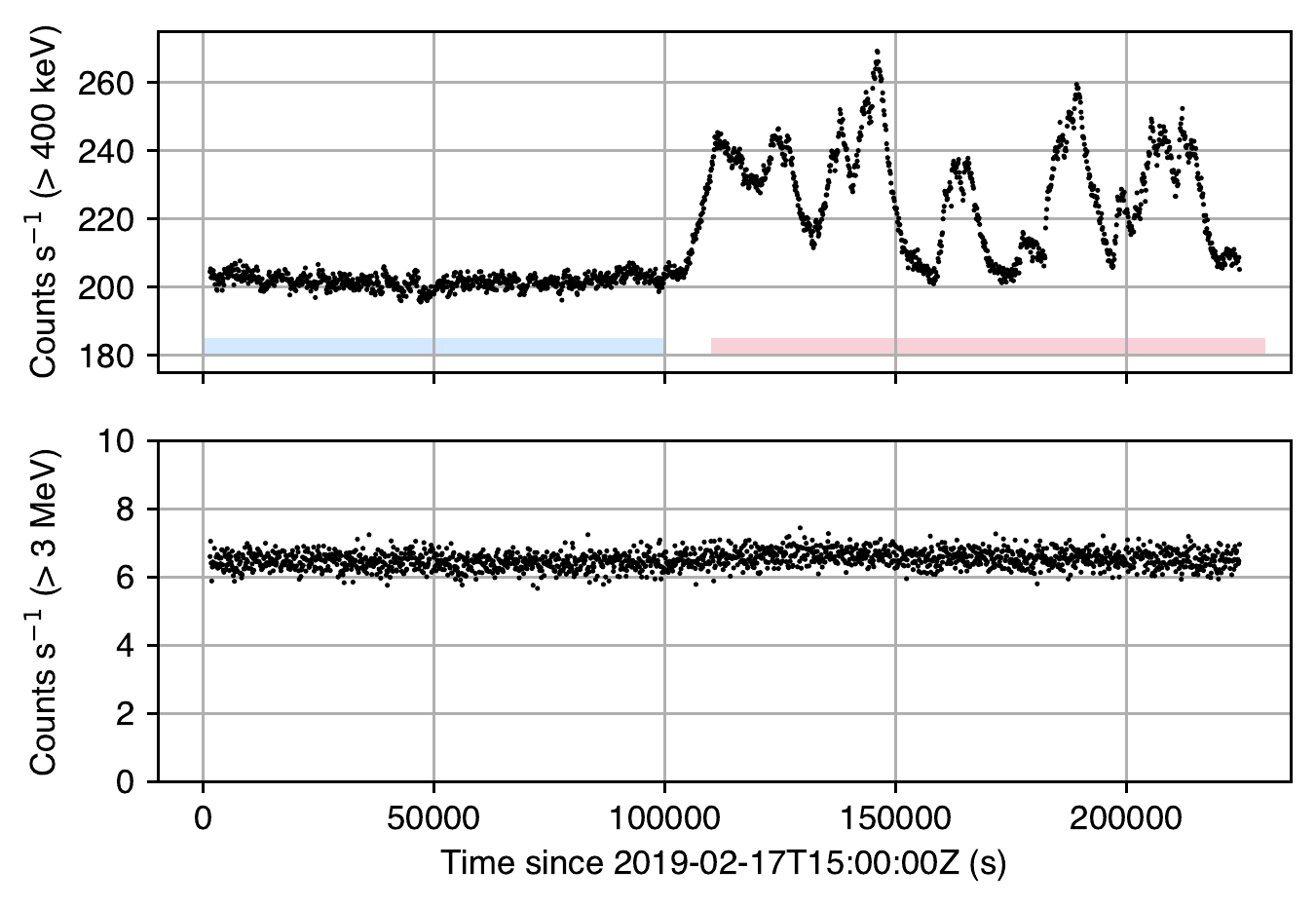}
\end{center}
\caption{Example count history of the environmental gamma-ray background recorded by one of the BGO-based detectors in Ishikawa in February, 2019 in the $> 400$~keV (top panel) and the $> 3$~MeV (bottom panel) energy bands. Three-day worth of data are shown. The count rate of the lower energy band (top panel) varies significantly after time $10^5$~s due to gamma rays from radioisotope washout due to precipitation, while that of the $> 3$~MeV band stays almost the constant. Blue and red rectangles indicate time periods of fair weather and intermittent rain, of which energy spectra are shown in Fig.~\ref{fig:background_spectra}. }
\label{fig:background_timehistory}
\end{figure}

\begin{figure}[htb]
\begin{center}
\includegraphics[width=0.9\hsize]{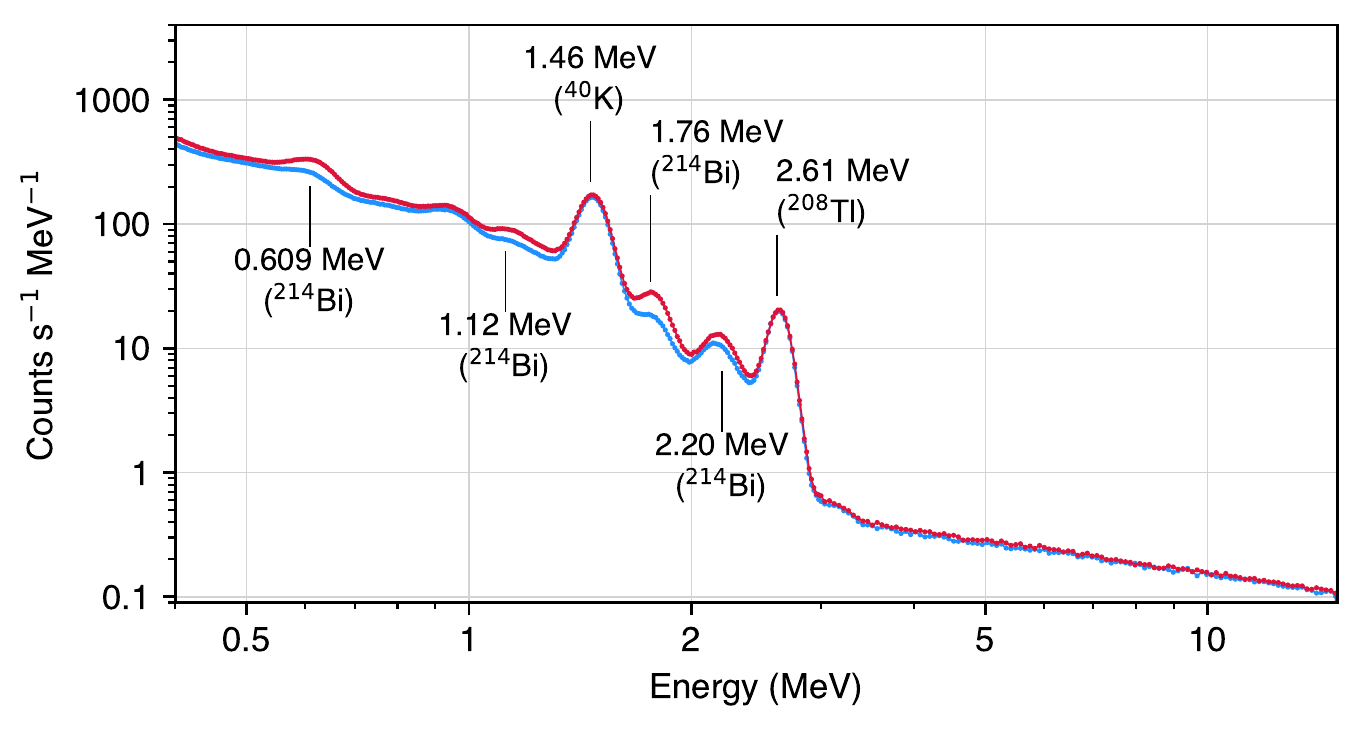}
\end{center}
\caption{Energy spectra of the environmental background extracted from the time periods without precipitation (blue) and with precipitation (red), as indicated in Fig.~\ref{fig:background_timehistory}. Statistical errors are plotted in the figure, but can be hardly visible due to high counting statistics.}
\label{fig:background_spectra}
\end{figure}

During the outdoor observation campaign, the temperature of the detector system (measured on the DSP board) varied between 25--60~$^{\circ}$C (the high temperature occurred under the clear skies, due to heating by the direct sun light).
Even with this temperature variation, energy scale did not change significantly; typical shifts of 1.46~MeV ($^{40}$K) and 2.61~MeV ($^{208}$Tl) peak centers in the ADC channel (i.e. raw voltage value before energy scale correction) were less than 3\% and 4\%, respectively. The 0.609~MeV line from $^{214}$Bi, which is clearly visible when there is precipitation, is used to validate the derived energy scale, and it was confirmed that the accuracy of the linear function is better than 2\% at 0.609 MeV for the BGO scintillation crystals. An example count history is shown in Fig.~\ref{fig:background_timehistory}, and spectra of the environmental background radiation during fair weather and precipitation are plotted in Fig.~\ref{fig:background_spectra}.

\subsection{Time assignment}\label{sec:timing_cal}
The analog daughter board carries a single-frequency GPS receiver with an external patch antenna. The navigation message output and the 1~pulse-per-second (PPS) signal of the module are routed to the FPGA on the main board via the stacking connector. On the rising edge of the 1~PPS signal, the FPGA logic registers the absolute time information in the navigation message along with a value of the free-running 48-bit time counter which is incremented at 100~MHz. The registered information is read by the DAQ software every 30~s, and stored in an output data file. The information is then used by the offline data processing pipeline to assign absolute time to each gamma-ray pulse which is recorded with a (free-running) time-counter value at trigger (i.e. when its pulse height exceeded the threshold value).

\begin{figure}[htb]
\begin{center}
\includegraphics[width=0.8\hsize]{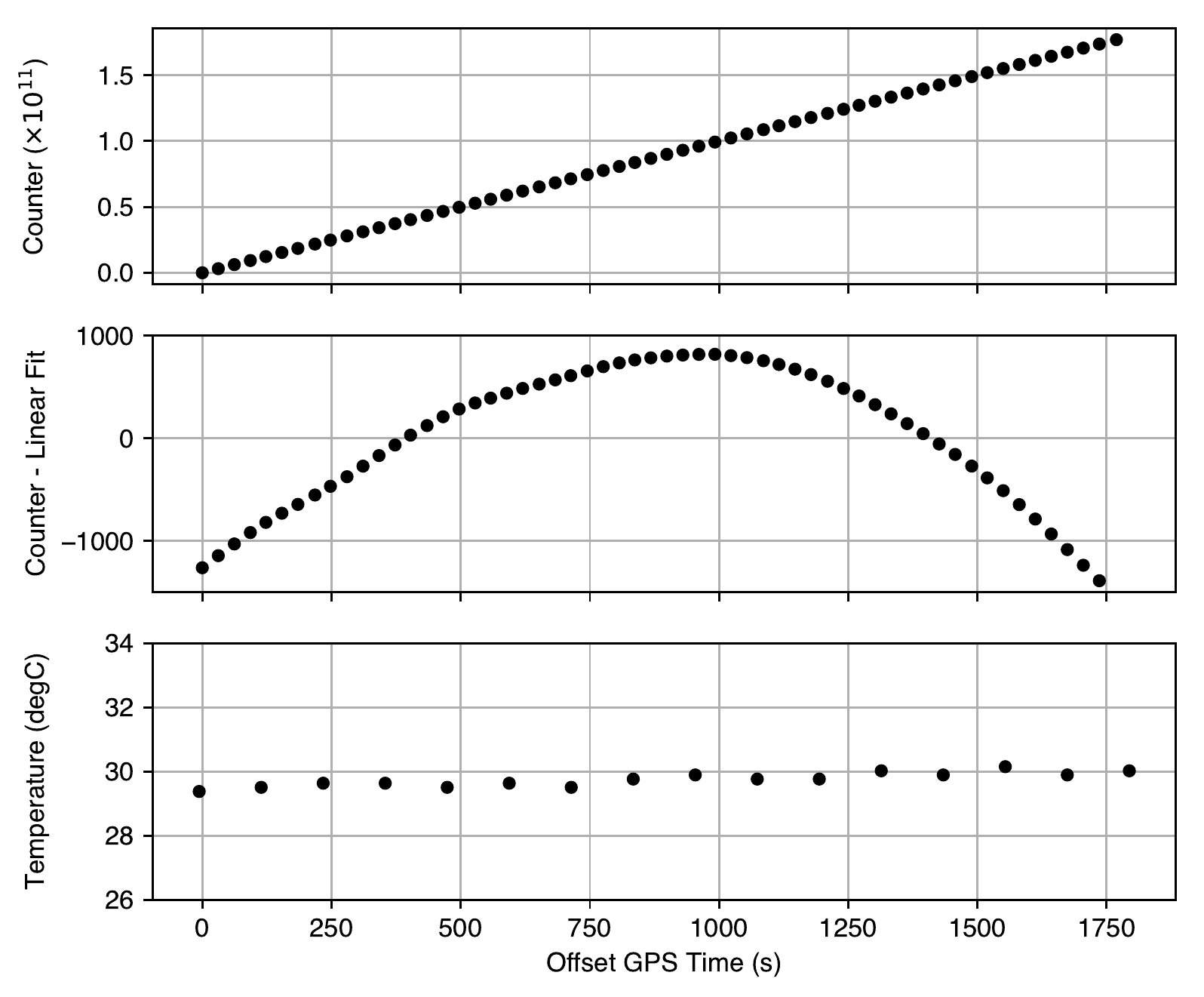}
\end{center}
\caption{Top: Typical relation of a counter incremented by the free-running 100-MHz clock fed to FPGA versus GPS time over 30~min (one observation interval).
Middle: The same counter value, but with the best-fit linear model being subtracted to visualize the time-variable drift of the local oscillator. A counter value of 1000 corresponds to 10~$\mu$s. 
Bottom: Temperature measured on the FPGA board.}
\label{fig:clock_fluctuation}
\end{figure}

Figure~\ref{fig:clock_fluctuation} shows an example of minute-scale variation of the local clock reconstructed based on the recorded GPS-based absolute time and the free-running time counter. When the receiver tracks sufficient number of GPS satellites, the accuracy of the 1~PPS signal generated by the module is reported to be $\sim$10~ns based on its data sheet. The sampling of ADC (50~MHz) governs the time resolution of trigger time of each gamma-ray pulse signal to be 20~ns. The time scale of scintillation photon emission (de-excitation) in the scintillation crystals ($\sim$a few hundred ns to 1~$\mu$s depending on crystals) and that of the band-pass-filtered pulse ($\sim$2~$\mu$s) are longer than the 1~PPS timing accuracy and the ADC sampling interval, and jitter of these components could potentially worsen the overall time accuracy. However, based on time correlation study between our gamma-ray measurement and radio-frequency observations (for example, \cite{Wada_2019_commphys}) confirmed that an absolute time accuracy better than 1~$\mu$s is achieved in this GPS-supported time assignment mode.

Occasionally, the GPS receiver did not generate navigation solution (thus no time information) due to low number of satellites in the field of view. In such a case, the pulse trigger time was converted to absolute time based on the system time of Raspberry Pi which was synchronized to the public NTP server via a cellular network. An absolute of time assignment in this mode is thought to be on the order of 10--100~ms, depending mostly on the round trip time of the cellular network.

\subsection{Search of TRB events}\label{sec:search_algorithm}
Gamma-ray enhancement events are usually found as an excess from the environmental background gamma-ray radiation, while the background itself is also variable.
As shown in Figures \ref{fig:background_timehistory} and \ref{fig:background_spectra}, background gamma-ray count rate ($\sim$6.5~counts~s$^{-1}$) varies significantly below 3~MeV depending on presence of precipitation, and this variation can lower sensitivity of the search. In contrast, the $>3$~MeV energy range is dominated by cosmic-ray induced signals, of which count rate is almost stable, and relatively lower than of the $<3$~MeV range.
Therefore, to lower the contamination from low-energy ($<3$~MeV) time-variable background signal, and to increase the signal-to-noise ratio, we implemented the following processes in the search algorithm;
1) count-rate history of photons with energies above 3~MeV is generated for each 30-min data chunk,
2) the 30-min count-rate history is further binned to a histogram, and a standard deviation is computed, 
3) the maximum count rate in the 30-min data chunk is divided by the standard deviation to derive ``significance'' value, after the mean count rate is subtract, and then,
4) a potential TRB event is reported when the ``significance'' exceeds a threshold value.
In the nominal batch analyses, we used a time bin width of 10~s and a significance threshold of 5~standard deviation; i.e. when a count-history bin contains gamma-ray counts which is more than 5~$\sigma$ apart (higher count rate) from the mean of the histogram, the bin is flagged for further examination by humans.

\begin{figure}[htb]
    \begin{center}
    \includegraphics[width=0.9\hsize]{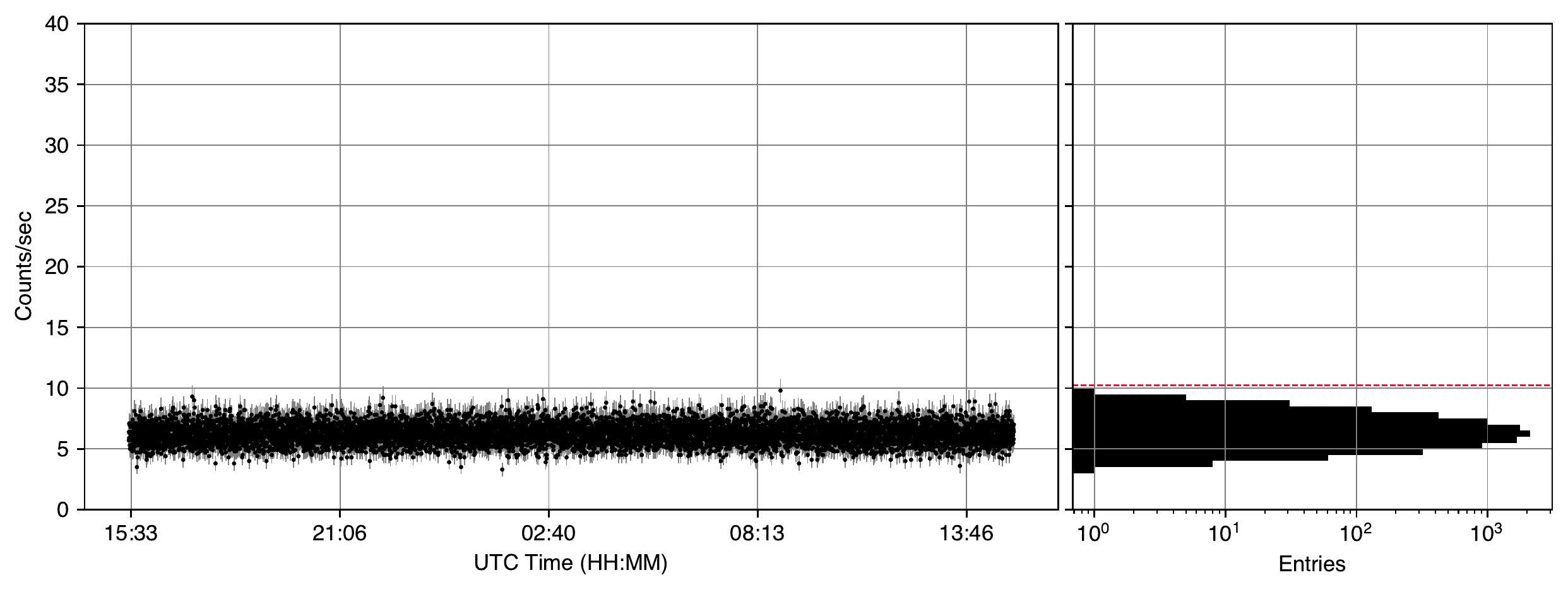}
    \includegraphics[width=0.9\hsize]{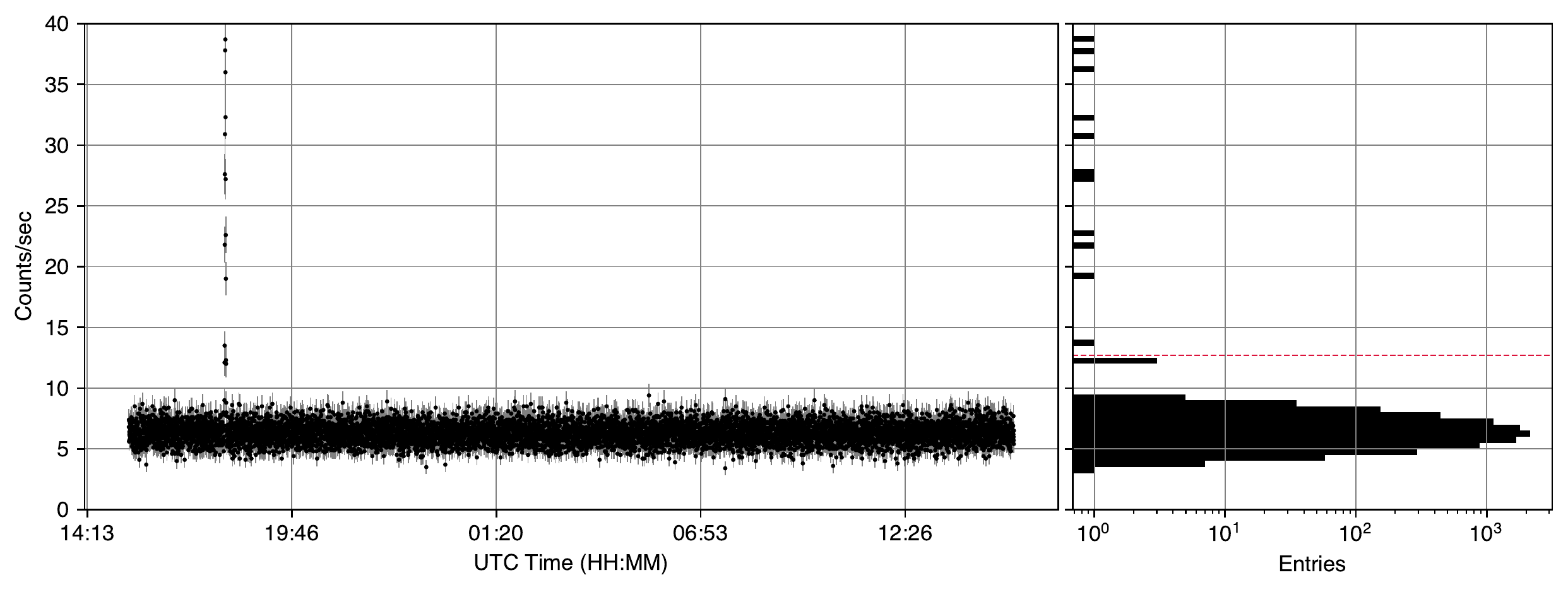}
    \end{center}
    \caption{
    Example of 24-hour count histories of photons with energies above 3~MeV, without (top panels) and with (bottom panels) count rate exceeding the event detection threshold. The histogram in the right panels show distribution of the count rate, with the event-detection threshold being shown as red dashed lines.
    Top and bottom rows show data of December 1st and December 6th, 2016, respectively, obtained by a detector deployed in Komatsu City, Ishikawa Prefecture.
    }
    \label{fig:trb_search}
\end{figure}

To illustrate this event search process, Fig.~\ref{fig:trb_search} presents two example 10-second-binned 30-min count histories, one with no significant count increase, and the other with a gamma-ray glow being detected.

\section{Results}

\subsection{Observation campaign}
\begin{figure}[htb]
\begin{center}
\includegraphics[width=0.7\hsize]{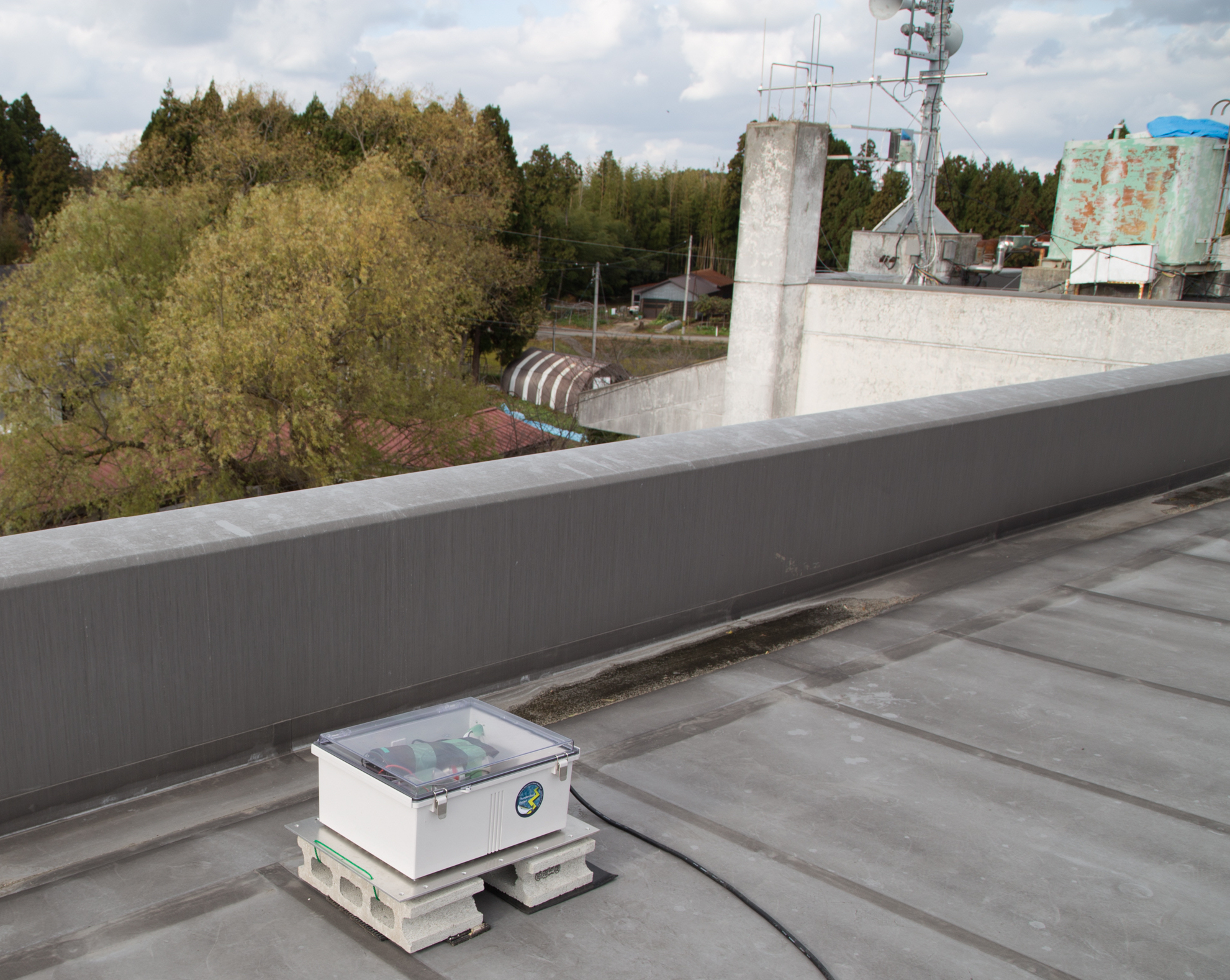}
\end{center}
\caption{Photograph of a detector system deployed on the roof of one of the observation sites.}
\label{fig:deployed_detector}
\end{figure}

\begin{figure}[htb]
    \begin{center}
    \includegraphics[width=0.7\hsize]{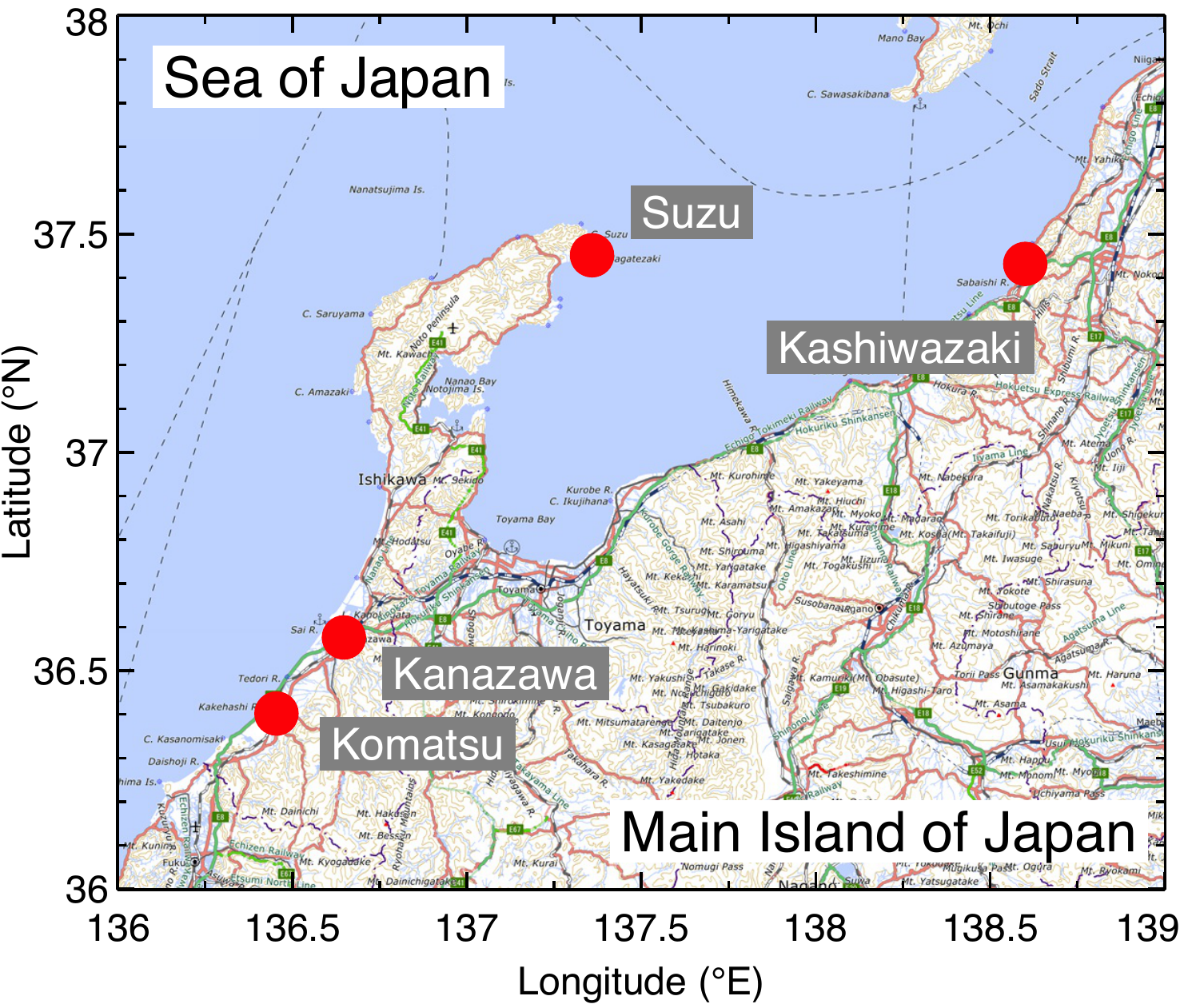}
    \end{center}
    \caption{
    Locations of the observation sites of the experiment.
    }
    \label{fig:install_map}
\end{figure}

\begin{figure}[htb]
    \begin{center}
    \includegraphics[width=0.9\hsize]{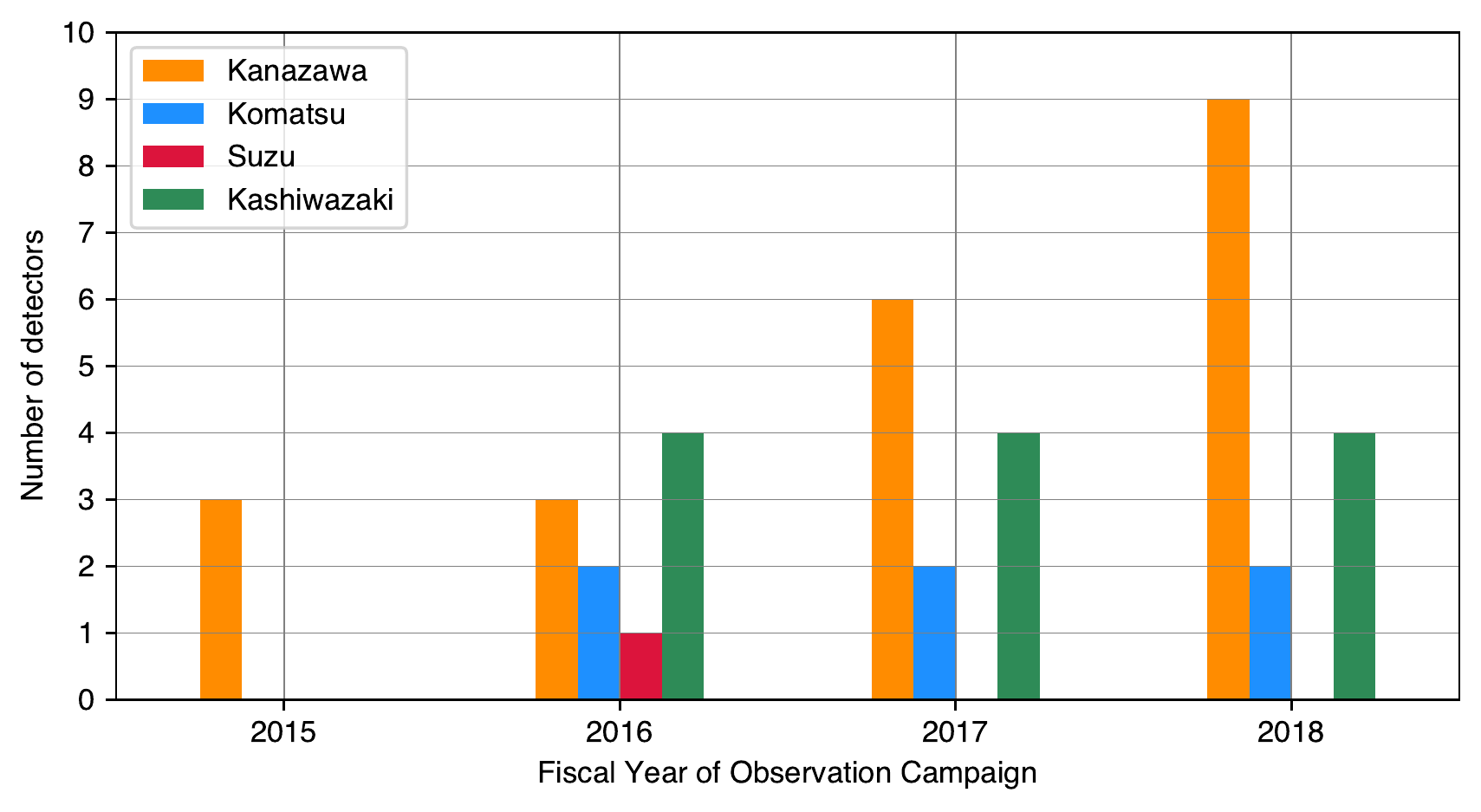}
    \end{center}
    \caption{Number of detectors deployed to each observation area.}
    \label{fig:deployment_hist}
\end{figure}

In 2015, we developed 4 prototype detectors, and started a multi-point observation campaign in Kanazawa City, Ishikawa Prefecture, with 3 detectors deployed in the city. The detectors were installed on the rooftop of a building of the observation sites, as exemplified in Fig.~\ref{fig:deployed_detector}.
In later years, we increased the number of detectors, and deployed in more observation sites in Ishikawa and Niigata Prefectures. Figure~\ref{fig:install_map} presents the locations of each observation site. Table \ref{tab:deployment_history} and Fig.~\ref{fig:deployment_hist} summarize the number of detectors that were deployed during annual observation campaigns since 2015. 
An annual observation campaign typically extends over 5 to 6 months from October or November to March next year; for example, the 2016 observation campaign started in November, 2016 and ended in March, 2017.

\begin{table}[!h]
\caption{The number of detectors deployed in each observation campaign, in each observation area, and the duration of each observation campaign in days.}
\label{tab:deployment_history}
\centering
\begin{tabular}{cccccc}
\hline
Prefecture & Area        & \multicolumn{4}{c}{Year}\\
           &             & 2015 & 2016 & 2017 & 2018\\
\hline
Ishikawa   & Kanazawa    &  3     & 3      & 6      & 9\\
           & Komatsu     &  0     & 2      & 2      & 2\\
           & Suzu        &  0     & 1      & 0      & 0\\
Niigata    & Kashiwazaki &  0     & 4      & 4      & 4\\
\hline
\multicolumn{2}{c}{Duration (days)} & 94 & 189 & 127 & 141\\
\multicolumn{2}{c}{Total (days)} & \multicolumn{4}{c}{551}\\
\hline
\end{tabular}\\
\end{table}

\subsection{Number of detected TRB events}

\begin{figure}[htb]
    \begin{center}
        \includegraphics[width=0.5\hsize]{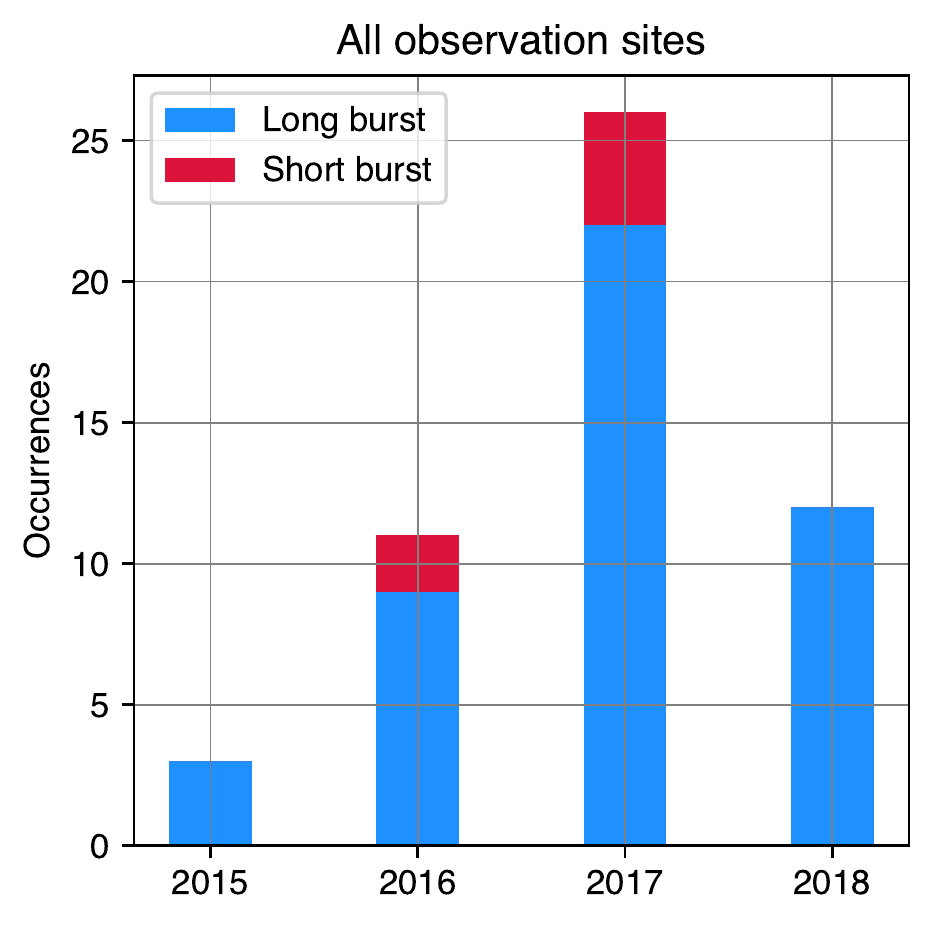}
        \includegraphics[width=1\hsize]{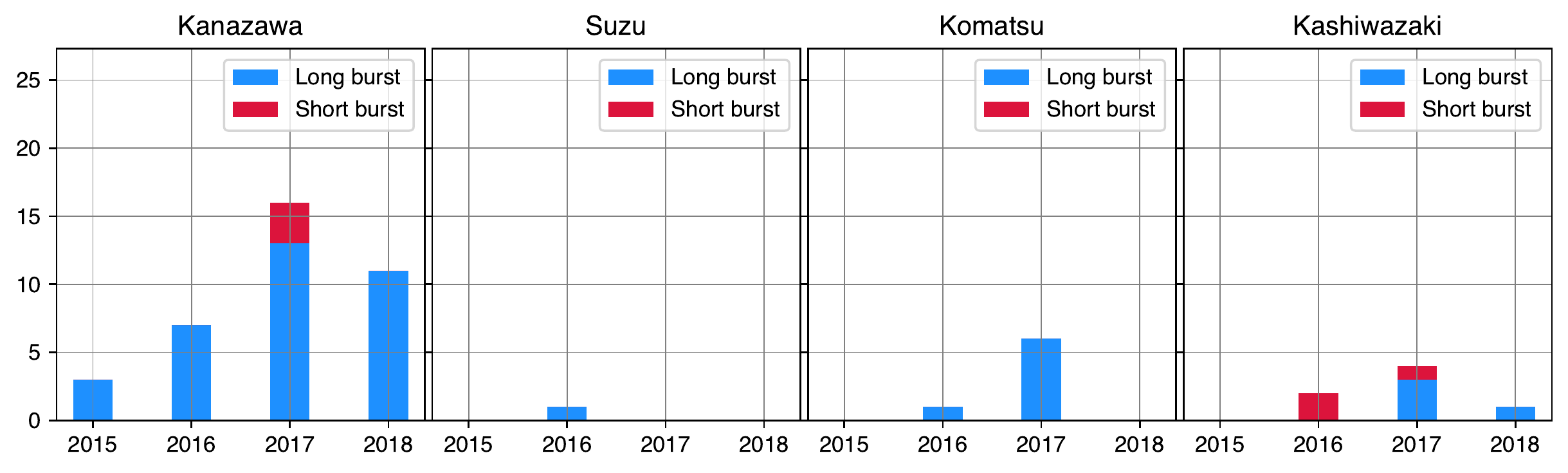}
    \end{center}
    \caption{
        Top panel: Number of TRB events detected in each observation campaign. Blue and red rectangles correspond to long- and short-duration TRBs, respectively.
        Bottom panels: The same as the top panel but events detected in each observation area are shown separately.}
    \label{fig:trb_hist}
\end{figure}

We have applied the event search algorithm described in Section \ref{sec:search_algorithm} to the data collected through the observation campaigns in the past 4 winter seasons (late 2015 to early 2019), and detected 46 long-duration bursts and 5 short-duration bursts. The two short-duration bursts detected in the 2017 campaign happened during simultaneously-observed long-duration bursts. 
Figure~\ref{fig:trb_hist} presents yearly histogram of the detected events.


\subsection{Multi-point detection of TRB}

The primary objective of the present experiment is to measure TRB events (both short-and long-duration gamma-ray bursts) with multiple detectors located in different sites, to study the physical extent of the gamma-ray emitting region in the cloud and its potential temporal/spatial variability as well as the movement of the cloud in detail. In fact, 14 events of all the detected TRB events were simultaneously observed by multiple detectors. 

\begin{figure}[htb]
    \begin{center}
        \includegraphics[width=\hsize]{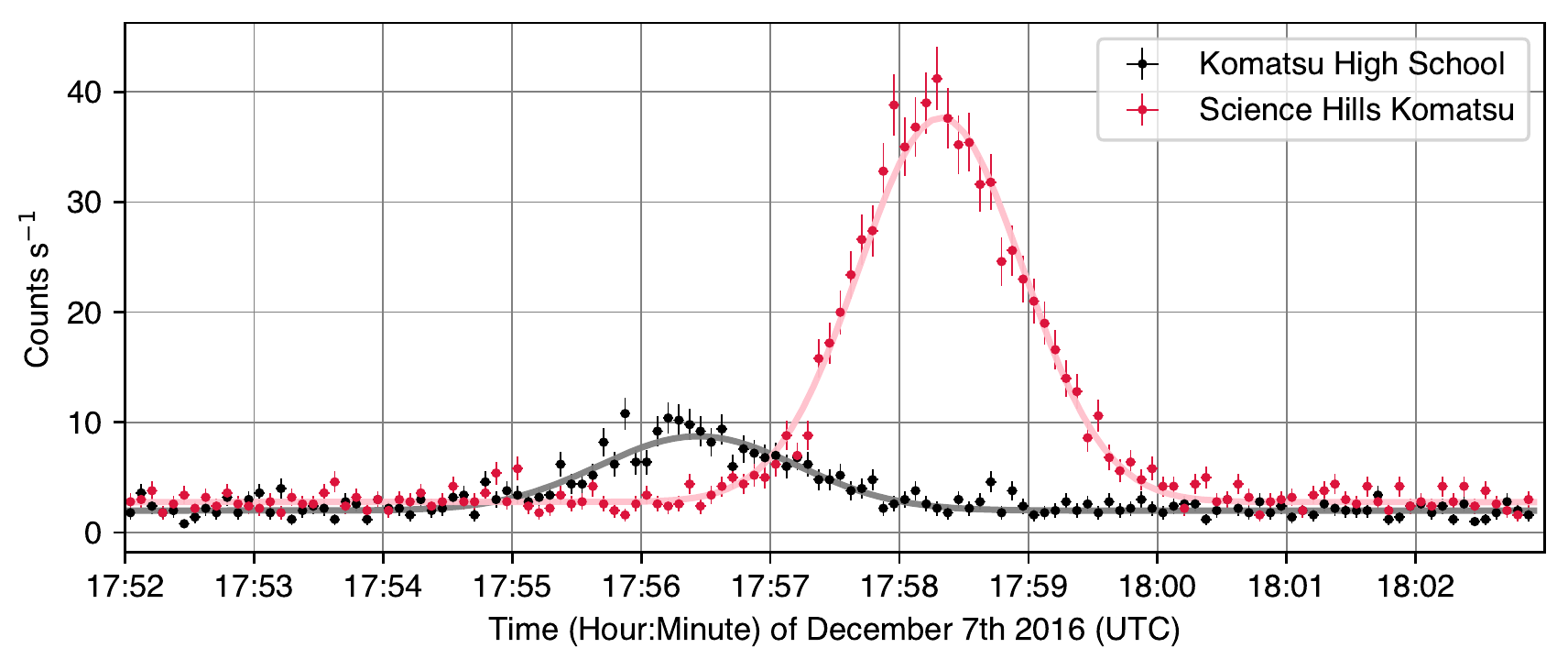}
    \end{center}
    \caption{
    Gamma-ray count-rate time histories recorded by our detectors in Komatsu High School (black filled circles) and Science Hills Komatsu (red filled circles) in the 3--15~MeV energy band, with a bin size of 10~s. Error bars show statistical errors.
    Solid black and red curves are the best-fit ``gaussian + constant'' model functions (Eq.~\ref{eq:count_history}).
    }
    \label{fig:komatsu_count_history}
\end{figure}

For example, Fig.~\ref{fig:komatsu_count_history} shows a gamma-ray glow event detected by two detectors in Komatsu City, Ishikawa Prefecture at $\sim$17:54:00$-$18:00:00 of December 7th, 2016 (UTC). In this event, the detector in Komatsu High School first detected enhanced gamma-ray counts starting at $\sim$17:54:00 and ending at $\sim$17:58:00. A minute later, at around 17:55:00, a similar count-rate increase was recorded by the detector in Science Hills Komatsu (the art science museum), and lasted till 18:00:00. These 3--15~MeV count-rate time profiles are well described by a gaussian function plus a constant (corresponding to background signal), 
\begin{equation}
f(t) = a\times\exp\left(-\frac{(t-b)^2}{c^2}\right) + d~[\mathrm{counts~s}^{-1}]\label{eq:count_history}
\end{equation}
where $t$ is time, $a$, $b$, $c$, and $d$ are a normalization factor, a peak-center time, a width of the gaussian, and the environmental background count rate, respectively. Table \ref{tab:komatsu_count_history} lists the best-fit parameters. The normalization factors and the widths yield the total counts of gamma rays from the gamma-ray glow, in 3--15~MeV, are estimated to be $\sim755\pm36$~counts and $\sim3310\pm58$~counts, in Komatsu High School and Science Hills Komatsu, respectively. The separation of the two gaussian centers is $114\pm3$~s. Errors are 1 standard deviation.

\begin{table}[!h]
\caption{Best-fit parameters of the count-rate time-profile model (Eq.~\ref{eq:count_history}). Errors are 1 standard deviation, and count rates are in the 3--15~MeV energy band.}
\label{tab:komatsu_count_history}
\centering
\begin{tabular}{cccccc}
\hline
Location & $a$ & $b$ & $c$ & $d$ & $\chi^2$~(n.d.o.f.)$^{1}$\\
        & counts~s$^{-1}$ & UTC & s & counts~s$^{-1}$ \\
\hline
Komatsu High School     & $6.7\pm0.4$ & 17:56:26$\pm4$  & $66\pm4$ & $2.0\pm0.1$ & 143.2 (133)\\
Science Hills Komatsu & $35.6\pm0.8$ &  17:58:19$\pm1$  & $52\pm1$ & $2.7\pm0.1$ & 148.8 (133)\\
\hline
\end{tabular}\\
$^{1}$ Chi-square value of the fit and the number of degrees of freedom in parentheses.
\end{table}

\begin{figure}[htb]
    \begin{center}
        \includegraphics[width=\hsize]{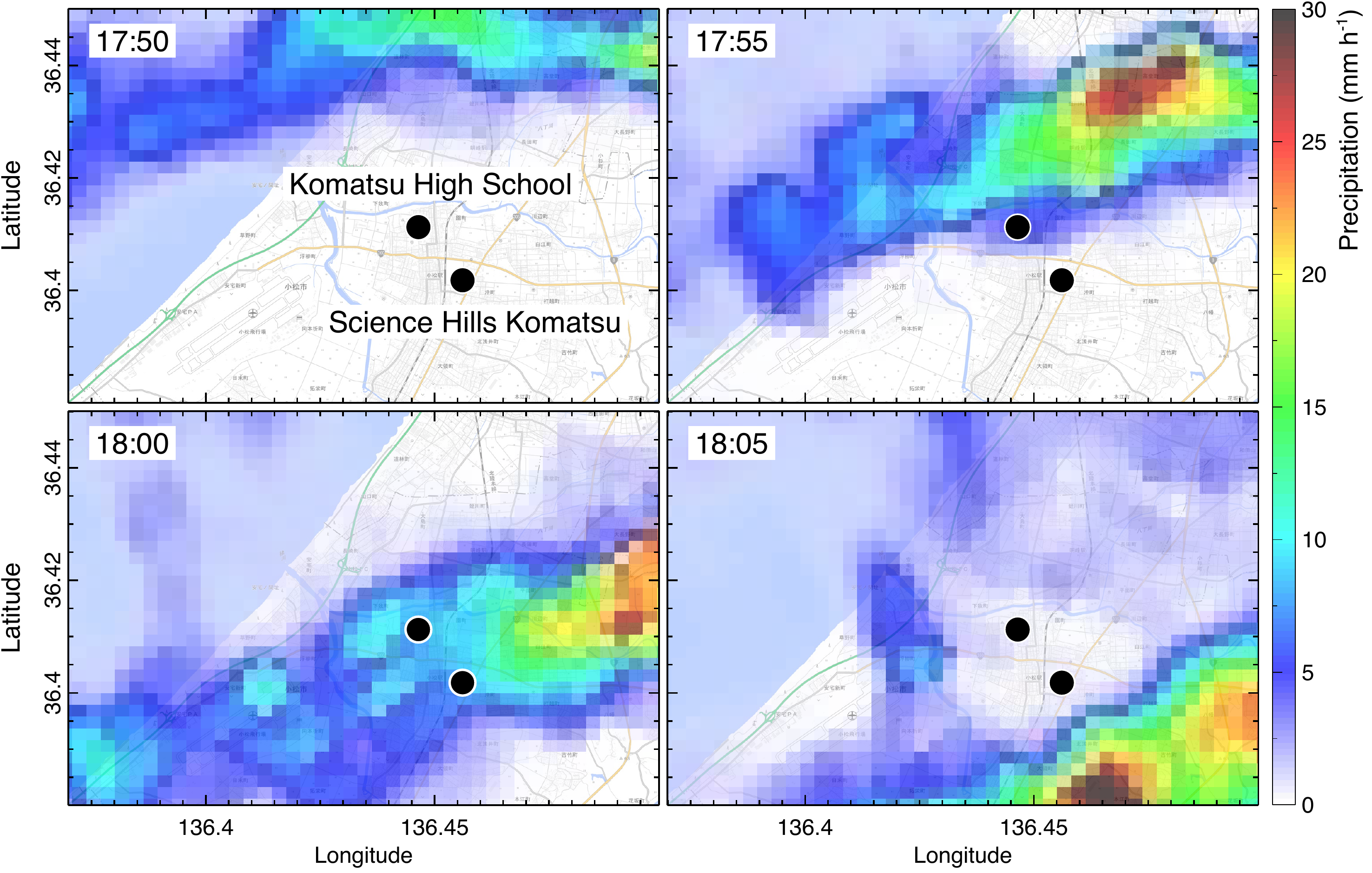}
    \end{center}
    \caption{
    The location of the two observation sites in Komatsu, Ishikawa Prefecture (filled circles).
    Precipitation intensity map obtained by XRAIN are shown for 4 five-minute intervals of December 7th, 2016 (UTC).
    }
    \label{fig:komatsu_radar}
\end{figure}

\begin{figure}[htb]
    \begin{center}
        \includegraphics[width=0.7\hsize]{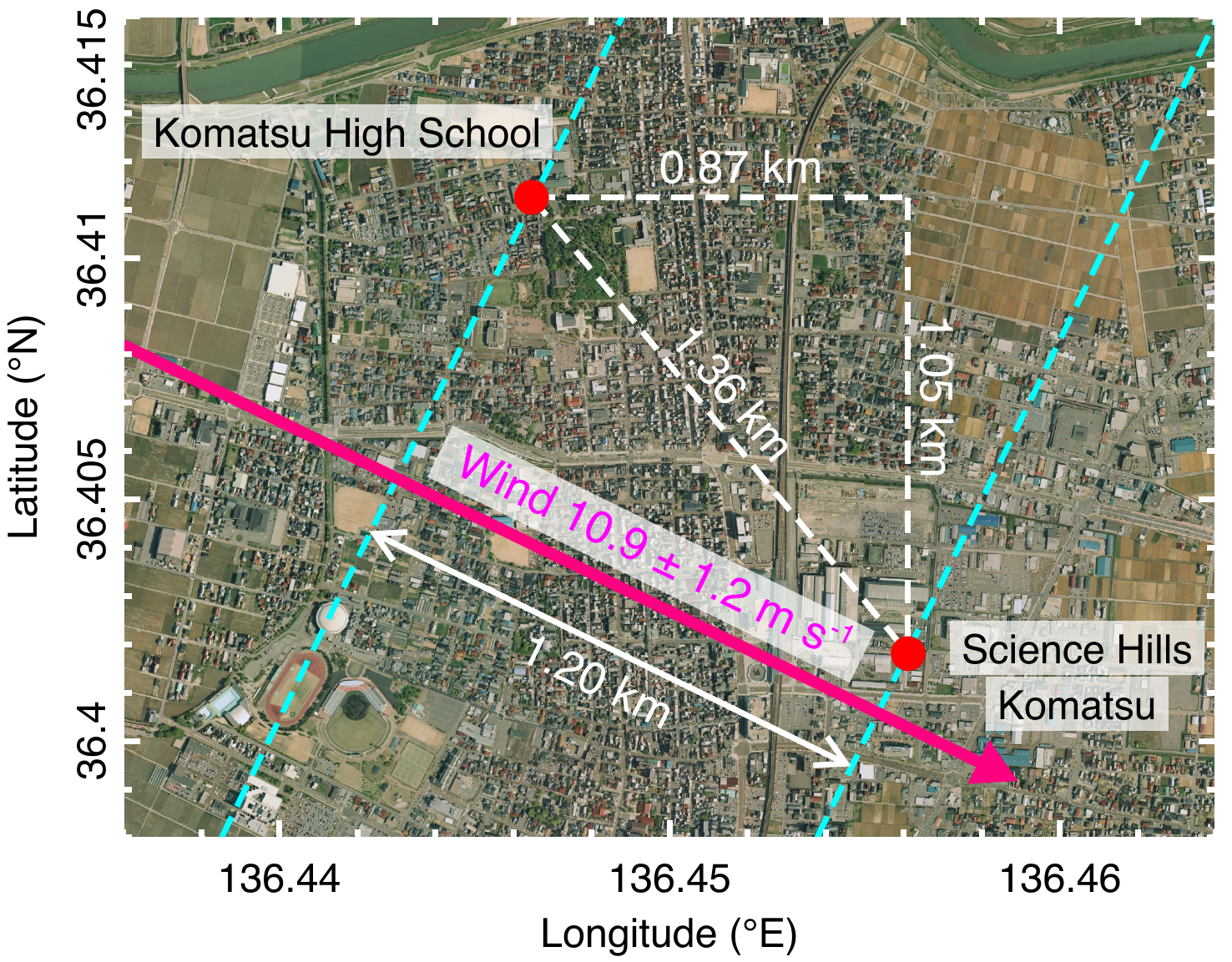}
    \end{center}
    \caption{
    Close-up view of the aerial photograph of Komatsu.
    Filled circles indicate the observation sites.
    Magenta arrow presents the wind direction estimated from the radar image analysis.
    White double-headed arrow is the hypothesized shortest distance as traveled by the brightest part of the gamma-ray emission region which yielded the highest peaks in the two count-rate histories. 
    }
    \label{fig:komatsu_map}
\end{figure}

The location of the two detectors deployed at Komatsu High School and Science Hills Komatsu are plotted in Fig.~\ref{fig:komatsu_radar}, with radar echo images taken during this time period being overlaid. The straight-line distance of the two sites is 1.36~km. By tracking the movement of the precipitation feature in the radar image, we estimated a wind speed of $10.9\pm1.2$~m~s$^{-1}$ and wind direction as shown in Fig.~\ref{fig:komatsu_map}. The wind direction is consistent with a hypothesis that a gamma-ray emitting region in the thundercloud was moving from west northwest to east southeast, first traveling over Komatsu High School, and arriving Science Hills Komatsu after that. Based on the estimated wind speed ($10.9\pm1.2$~m~s$^{-1}$) and the distance measured along the wind (1.20~km), a hypothetical travel time of the gamma-ray emission region can be estimated to be $110\pm12$~s. This value is consistent within errors with the peak-time difference based on the gaussian fitting ($114\pm3$~s), and therefore we consider that the wind speed and direction estimated based on the radar images are sufficiently accurate to be used in interpreting the temporal and the geometrical aspects of this particular gamma-ray glow event.

As mentioned above, the total gamma-ray count of Science Hills Komatsu is larger than that of Komatsu High School by a factor of 4.4. Based on this combined with the wind direction, we infer that Science Hills Komatsu was (laterally) closer to the electron acceleration region in the thundercloud, and observed less attenuated gamma rays than the other.

\begin{figure}[htb]
    \begin{center}
        \includegraphics[width=0.8\hsize]{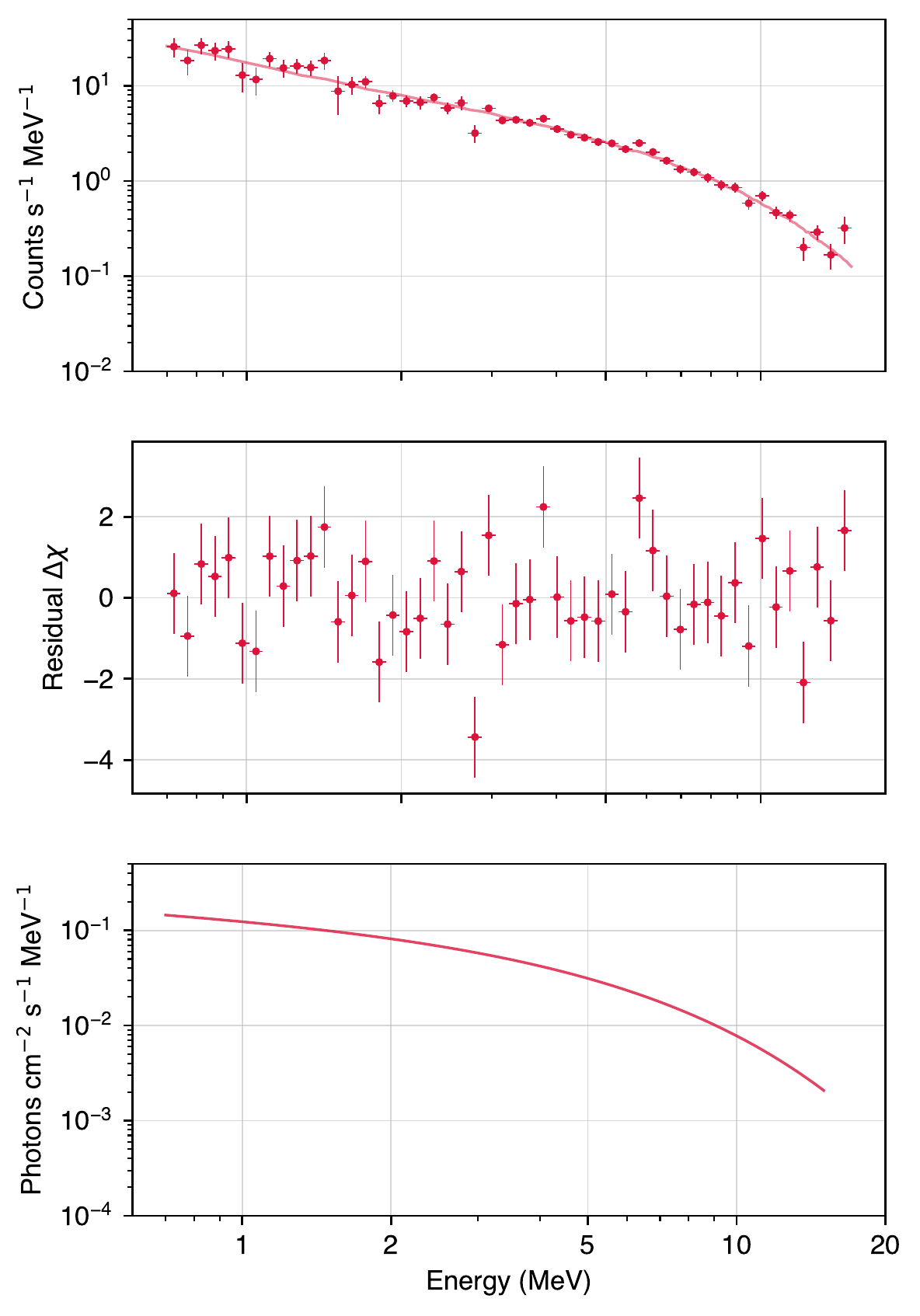}
    \end{center}
    \caption{
	Top panel: Gamma-ray energy spectrum of the gamma-ray glow event recorded at Science Hills Komatsu (red filled circles). Solid red curve is the best-fit power-law with exponential cutoff (Eq.~\ref{eq:cutoffpl}). The model is convolved with the detector's energy response function so that the fitting is performed in the detector count-rate dimension.
	Middle panel: Fit residual computed as (data $-$ model)/error.
	Bottom panel: The same best-fit model function as that of the top panel, but without being convolved with the detector energy response function. Note that ordinate is in units of photons~cm$^{-2}$~s$^{-1}$~MeV$^{-1}$, which represents the photon flux arriving at the detector.
    }
    \label{fig:komatsu_spectrum}
\end{figure}

The high counting statistics of the Science Hills Komatsu data allowed us to extract an energy spectrum of the gamma-ray glow event, as shown in Fig.~\ref{fig:komatsu_spectrum}. The spectrum of the glow event was extracted from a time range 17:56:30--18:00:00 (UTC). The environmental background signals were extracted using two 60-s chunks of data before and after the glow event, and subtracted from that of the glow event. Based on previous spectral studies \cite{Tsuchiya_2011}, we tried to characterize the spectral shape by fitting it with a power law with an exponential cutoff:

\begin{equation}
f(E) = N \times E^{-\Gamma} \exp(-E/E_\mathrm{c})\label{eq:cutoffpl}~\mathrm{photons}~\mathrm{cm}^{-2}~\mathrm{s}^{-1}~\mathrm{MeV}^{-1}
\end{equation}
where $E$ is gamma-ray energy in MeV, and $N$, $\Gamma$, and $E_\mathrm{c}$ are a normalization factor in
photons~cm$^{-2}$~s$^{-1}$~MeV$^{-1}$, a power-law photon index, and
a scaling factor for the exponential cutoff, respectively. An energy response function of the detector was generated based on a Monte-Carlo simulation using the particle transport framework \texttt{Geant4} \cite{Agostinelli_2003,Allison_2006,Allison_2016}, and was convolved with the model function during the fitting which happened in the detector count-rate dimension. The $\chi^2$ value, which was computed as a square sum of difference between the model and the data divided by the statistical error, was minimized using the Levenberg-Marquardt algorithm in the SciPy software package. With the best-fit model parameters listed in Table \ref{tab:komatsu_spectrum}, the model reproduced the data reasonably well with no particular structure in the fit residual (middle panel of Fig.~\ref{fig:komatsu_spectrum}), with a null hypothesis probability of 7.5\%. When the same spectrum was fitted with a simple power law, a significant ``convex''-shaped systematic residual was seen with a large (unacceptable) $\chi^2$ value of 450 for 47~degrees of freedom, supporting the presence of a spectral cutoff feature. Because the electron bremssstrahlung is thought to be the primary emission process in the gamma-ray glow, the cutoff energy should have a close relation with the maximum energy of accelerated electrons (e.g. \cite{Dwyer_2012a} for a detailed Monte-Carlo simulation study of the acceleration and the emission processes). In addition, we anticipate that statistical analyses of the spectral shape and their temporal evolution based on multi-point observation data will allow us to better constrain the properties of the electron acceleration (electric field strength, lateral extent of the acceleration region), and plan to publish a consolidated result elsewhere.

\begin{table}[!h]
\caption{Result of energy-spectral model fitting with a power law with an exponential cutoff (Eq.~\ref{eq:cutoffpl}) to the Science Hills Komatsu data. Errors are at the 90\% confidence level.}
\label{tab:komatsu_spectrum}
\centering
\begin{tabular}{cccccc}
\hline
N & $\Gamma$ & $E_\mathrm{c}$ & Energy flux$^{1}$ & $\chi^2$~(n.d.o.f.)$^{2}$ \\
ph~cm$^{-2}$~s$^{-1}$~MeV$^{-1}$ &   & MeV & MeV~cm$^{-2}$~s$^{-1}$ & \\
\hline
$0.158^{+0.015}_{-0.016}$ & $0.26^{+0.14}_{-0.15}$ & $4.10^{+0.51}_{-0.33}$ &
1.18 & 60.5 (46)\\
\hline
\end{tabular}\\
$^{1}$ Energy flux in the 3--15~MeV energy band.\\
$^{2}$ Chi-square value of the fit and the number of degrees of freedom in parentheses.\\
\end{table}

\clearpage
\section{Science highlights}

In this section, we review new findings and advancements of our understanding on high-energy radiation from lightning and thundercloud based on our publications which utilized data collected with our detector system.

\subsection{Photonuclear reaction triggered by a downward TGF}
Enoto et al. \cite{Enoto_2017} reported a sub-millisecond intense gamma-ray flash (downward TGF) and a subsequent short-duration gamma-ray burst lasting for $\sim200$~ms, recorded on February 6th, 2017, at our observation site in Kashiwazaki-Kariwa nuclear power station in Niigata Prefecture.

\begin{figure}[htb]
\begin{center}
\includegraphics[width=0.9\hsize]{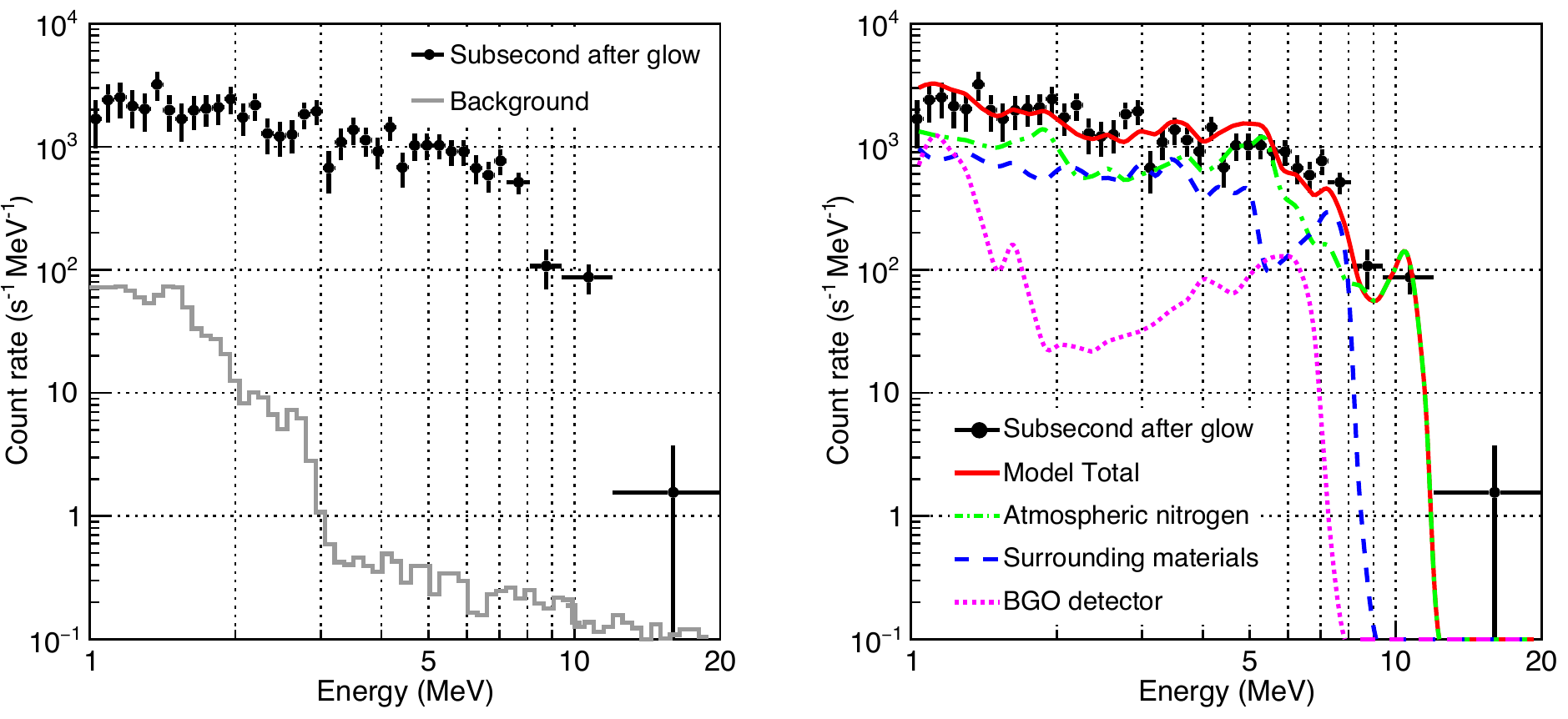}
\end{center}
\caption{Observed gamma-ray spectra of the short-duration gamma-ray burst (black filled circle) and model spectrum constructed based on the Monte-Carlo simulation of neutron-induced nuclear gamma rays (green dash-dotted, blue dashed, and purple dotted curves). Red solid curve shows the sum of the individual model components.}
\label{fig:enotoeal_spectra}
\end{figure}

As shown in Fig. \ref{fig:enotoeal_spectra}, the energy spectrum of the short-duration burst consisted of an extremely ``flat'' or ``hard'' continuum (a photon indices $\Gamma\sim0.5$ when fitted with a power-law function of $N \times (E/1~\mathrm{MeV})^{-\Gamma}$ where $N$ and $E$ are a normalization factor and gamma-ray energy), associated with an abrupt cutoff at $\sim$10~MeV. These features made the spectrum look very different from typical energy spectra of bremssstrahlung emission seen e.g. in typical gamma-ray glows (e.g. Fig.~\ref{fig:komatsu_spectrum}).

The short-duration burst was followed by a minute-lasting gamma-ray burst. The energy spectrum of this distinctive emission, in turn, predominantly consisted of electron-positron annihilation gamma-ray line at $511$~keV and its Compton scattered continuum signals.

After extensive spectral, temporal, and simulation studies, we showed unequivocally that a lightning discharge emitted huge amount of energetic ($>10$~MeV) gamma rays, and neutrons were produced via atmospheric photonuclear reactions
(such as $\gamma + ^{14}$N $\to$ $^{13}$N$+$n).
The short-duration burst was interpreted well as a superposition of nuclear gamma-ray lines emitted from nuclei that underwent neutron capture, and the peculiar minute-lasting annihilation gamma-ray line emission was explained as a result of $\beta+$ decay of unstable nuclei (again, produced via photonuclear reaction).

Production of neutrons via the photonuclear reaction has been suggested based on observational results \cite{Gurevich_2012,Chilingarian_2012b,Bowers_2017}, and theoretical studies \cite{Babich_2006,Babich_2007,Carlson_2010}, there have been multiple reports on potential detection of neutron signals from thundercloud- and lightning-related high energy radiation (for complete reference list, see \cite{Enoto_2017}). Our observation provided multi-point time-resolved data that confirm a) intense gamma-ray flash that caused neutron via photonuclear reaction, b) presence of unbound neutrons (via gamma-ray lines from neutron capture), and c) 511~keV annihilation lines from $\beta^{+}$-decay radioisotopes generated by photonuclear reaction. These formed the first comprehensive observational evidence of such an exotic photonuclear reaction happening in the Earth's dense atmosphere.

\subsection{Physical properties of downward TGF}
On November 24th, 2017, three of our detectors deployed in Niigata, Japan, detected four bunches of intense short-duration ($\ll1$~ms) gamma-ray flashes (TGFs), followed by exponentially-decaying $\sim200$-ms signals, which is, again, considered to be a result of photonuclear reaction (gamma-ray signals from de-excitation of isotopes generated via neutron capture).

We analysed time-resolved gamma-ray signals from our detectors, integrated radiation dose measured by argon ionization chambers, and low-frequency radio (LF) observations.
Our scintillation-crystal based detectors were heavily saturated by the intense gamma-ray signals from the four pulses of the downward TGFs, and therefore could not provide the total number of gamma-rays that entered the detector nor spectral information of the TGF.
Though, we were able to derive arrival times of the four TGF events with an accuracy of $\sim200$~$\mu$s. Comparison of these TGF event times  against LF time-series data showed clear correlation between TGFs and positive unipolar pulses (first and second gamma-ray flashes) or bipolar pulses (third and forth ones).

Compared to a scintillation-crystal based photon counter, an ionization chamber is much tolerant to high-flux radiation when effective area is approximately the same because the latter measures the amount of integrated ionization at the cost of fine time and energy resolutions. In the TGF event in question, the ionization chambers successfully provided accurate dose information at 5 locations (400--1900~m horizontally from the estimated location of TGF), as anticipated.
These dose data, combined with a Monte-Carlo simulation of gamma-ray emission and propagation in the atmosphere, were used to estimate an altitude of electron acceleration to be 2.5$\pm$0.5~km from the sea level.
Based on the altitude and the measured radiation dose, the total number of avalanche electrons ($>1$~MeV) was computed to be $8^{+8}_{-4}\times10^{18}$, which is approximately in the same range as those of accelerated electrons estimated from space-based observations of upward TGFs ($4\times10^{16}$--$3\times10^{19}$ by \cite{Mailyan_2016}), while many of TGFs observed in space are thought to originate at altitudes higher than 8~km \cite{Cummer_2015,Mailyan_2016}.

\subsection{The end of gamma-ray glow from thundercloud}

One of key questions the GROWTH project is set to answer is how stable electron-accelerating region starts to form, evolves over time, and disappears in thundercloud, or in other words, the life cycle of the source of gamma-ray glow. When the close phase of the life cycle is concerned, multiple previous measurements reported abrupt termination of gamma-ray glow that coincided with lightning discharge (\cite{Tsuchiya_2013,Chilingarian_2017,Chilingarian_2020} and references therein).

For revealing precise relationship between lightning discharge and cessation of gamma-ray glow, Wada et al. \cite{Wada_2019_prl} analysed an abrupt-termination event that was observed in Ishikawa Prefecture, on February 11th, 2017, by combining gamma-ray data collected by our detector and one from the GODOT project \cite{Bowers_2017} as well as LF data collected by multiple receivers located $\sim50$~km from the gamma-ray observation site. Although there have been previous reports of abruptly-terminated gamma-ray glows coinciding with radio frequency observations of lightning discharges that triggered the termination \cite{Chilingarian_2017}, the nature of single-site measurements of radio signal did not allow a detailed position and time correlation study between gamma-ray glows and lightning discharges.

However, in our study \cite{Wada_2018}, a multi-site LF observation provided, for the first time, a fine time- and position-resolved structure of an intercloud/intracloud lightning discharge that coincided with the gamma-ray glow termination and extended over $\sim60$~km lateral area with a 300~ms duration. Time association with the LF data and the gamma-ray data revealed that the termination happened when a stepped leader of the lightning discharge passed over the gamma-ray observation site with a horizontal distance of $0.7$~km. Since the discharge started prior to the abrupt termination of gamma ray about 15~km away from the gamma-ray observation site, causality in this event is obvious; lightning discharge affected the electric field structure and effectively disabled acceleration. Still, due to the long distance between the event and the LF observation sites ($\sim50$~km), we were unable to resolve vertical structure of the discharge. Continued simultaneous observation in gamma ray and radio frequency is anticipated to shed light on the charge structure in the cloud in such events in the future.

\section{Conclusion}

\begin{itemize}
\item Aiming at multi-point observation of particle acceleration and high-energy gamma-ray emission of thundercloud and lightning, we launched a new experiment campaign called ``Thundercloud Project'' in 2015, and developed a new, compact gamma-ray detector system (35$\times$45$\times$20~cm$^3$ in size) each carrying BGO or CsI scintillation crystal.
\item We have deployed 15 detectors to four cities in Ishikawa Prefecture and Niigata Prefecture in Japan in four winter seasons in 2015--2019, and accumulated 46 long-duration and 5 short-duration gamma-ray burst events, respectively.
\item Some of these events, for example the short-duration burst on February 6th, 2017 in Niigata, allowed us to record the whole process of downward TGF followed by photonuclear reaction and a traveling positron-emitting isotope cloud.
\item On long-duration burst, we have revealed that the long-duration gamma-ray burst can be abruptly terminated by a passage of a developing lightning leader (separated by 700~m horizontally) based on February 11th, 2017 data collected in Ishikawa Prefecture \cite{Wada_2018}. This is another stepping stone for understanding the life cycle of particle acceleration region in a thundercloud.
\item With accurate timing information with GPS, we have been able to correlate our gamma-ray data with radio frequency observations, enabling multi-messenger studies of high-energy activities of thundercloud and lightning.
\item We will continue observation campaigns in coming winter seasons.
\end{itemize}

\section*{Acknowledgment}

We deeply thank
D.~Yonetoku and T.~Sawano (Kanazawa University),
K.~Watarai (Kanazawa University High School),
K.~Yoneguchi and Kanazawa Izumigaoka High School, 
K.~Kimura (Komatsu High School),
K.~Kitano (Science Hills Komatsu),
K.~Kono (Ishikawa Plating Industry Co., Ltd),
Kanazawa Nishi High School,
Industrial Research Institute of Ishiwaka,
Sakaida Fruits,
Kanazawa Institute of Technology, 
Television Kanazawa Corporation,
Ishikawa Prefectural University, 
Ishikawa Prefectural Institute of Public Health and Environmental Science,
Kanazawa University Noto School, 
Norikura Observatory of Institute of Cosmic-Ray Research, The University of Tokyo,
Non-Profit Organization Mount Fuji Research Station,
Niigata Prefectural Radiation Monitoring Center, and
the Radiation Safety Group of the Kashiwazaki-Kariwa nuclear power station, Tokyo Electric Power Company Holdings, Incorporated for the support of detector deployment.
We are grateful to H.~Sakurai, M.~Niikura and the Sakurai group members at Graduate School of Science, The University of Tokyo for providing the BGO scintillation crystals.
T.~Nakano, T.~Tamagawa (RIKEN), A.~Bamba, H.~Odaka (The University of Tokyo),
D.~Umemoto (Kobe University), 
M.~Sato and Y.~Sato (Hokkaido University),
H.~Nanto, G.~Okada (Kanazawa Institute of Technology) 
M.~Kamogawa (University of Shizuoka),
G.~S.~Bowers (Los Alamos National Laboratory),
D.~M.~Smith~(University of California Santa Cruz), 
T.~Morimoto (Kindai University),
Y.~Nakamura (Kobe City College of Technology),
A.~Matsuki and M.~Kubo (Kanazawa University),
T.~Ushio (Osaka University), and
H.~Sakai (University of Toyama) contributed to the project through detector development, data provisioning, and discussions on results.
S.~Otsuka (The University of Tokyo) and H.~Kato (RIKEN) also supported detector developments.

The Monte-Carlo simulations were performed on the HOKUSAI GreatWave and BigWaterfall supercomputing systems operated by RIKEN Advanced Center for Computing and Communication.
The maps and aerial photos used in the figures are taken from Geospatial Information Authority of Japan. Data of XRAIN are provided by the Ministry of Land, Infrastructure, Transport and Tourism  via Data Integration and Analysis System (DIAS), The University of Tokyo.

This project is supported by
RIKEN Special Postdoctoral Researchers Program,
JSPS/MEXT KAKENHI grants (15K05115, 15H03653, 16H04055, 16H06006, 16K05555, 17K12966, 18J13355, 19H00683),
Hakubi project and SPIRITS 2017 of Kyoto University, and 
the joint research program of the Institute for Cosmic Ray Research (ICRR), The University of Tokyo.

The bootstrapping phase of this project was supported by a crowdfunding campaign named Thundercloud Project on the academic-crowdfunding platform ``academist''.
We thank Y.~Shikano, Y.~Araki, M.~T.~Hayashi, N.~Matsumoto, T.~Enoto, K.~Hayashi, S.~Koga, T.~Hamaji, Y.~Torisawa, S.~Sawamura, J.~Purser, S.~Suehiro, S.~Nakane, M.~Konishi, H.~Takami, T.~Sawara, and all other supporters of the crowdfunding campaign, and Adachi design laboratory for the support of creation of digital contents and visuals.

\bibliographystyle{ptephy}

\end{document}